\documentclass[trackchanges]{aastex701}

\usepackage{booktabs}        
\usepackage{xcolor}          
\usepackage{array}           
\usepackage[table]{xcolor}   

\begin{document}

\title{The 256-antenna Coherent All-Sky Monitor}

\author[0000-0003-1183-9293]{Liam Connor}
\affiliation{Center for Astrophysics $|$ Harvard \& Smithsonian, 60 Garden Street, Cambridge, MA 02138, USA}
\email{liam.connor@cfa.harvard.edu}

\author[0000-0002-7252-5485]{Vikram Ravi}
\affiliation{Cahill Center for Astronomy and Astrophysics, California Institute of Technology, Pasadena, CA 91125, USA}
\email{}

\author[0000-0001-5504-229X]{Pranav Sanghavi}
\affiliation{Center for Astrophysics $|$ Harvard \& Smithsonian, 60 Garden Street, Cambridge, MA 02138, USA}
\email{}

\author[0000-0003-3244-2711]{Vishnu Balakrishan}
\affiliation{Center for Astrophysics $|$ Harvard \& Smithsonian, 60 Garden Street, Cambridge, MA 02138, USA}
\email{}

\author{Luke Chung}
\affiliation{Center for Astrophysics $|$ Harvard \& Smithsonian, 60 Garden Street, Cambridge, MA 02138, USA}
\email{}

\author{Saren Daghlian}
\affiliation{Cahill Center for Astronomy and Astrophysics, California Institute of Technology, Pasadena, CA 91125, USA}
\email{}

\author[0000-0003-1183-9293]{Liam Dunn}
\affiliation{Fourier Space}
\email{}

\author[0000-0001-5169-2105]{Anthony Griffin}
\affiliation{Fourier Space}
\email{}

\author{Charlie Harnach}
\affiliation{Cahill Center for Astronomy and Astrophysics, California Institute of Technology, Pasadena, CA 91125, USA}
\email{}

\author{Mark Hodges}
\affiliation{Cahill Center for Astronomy and Astrophysics, California Institute of Technology, Pasadena, CA 91125, USA}
\email{}

\author{Andrew Jameson}
\affiliation{Fourier Space}
\email{}

\author{Michael Gutierrez}
\affiliation{Cahill Center for Astronomy and Astrophysics, California Institute of Technology, Pasadena, CA 91125, USA}
\email{}

\author[0000-0002-4209-7408]{Calvin Leung}
\affiliation{Department of Astronomy, University of California, Berkeley, CA 94720, USA}
\email{}

\author{Mei Lin}
\affiliation{Center for Astrophysics $|$ Harvard \& Smithsonian, 60 Garden Street, Cambridge, MA 02138, USA}
\email{}

\author{Advait Mehla}
\affiliation{Cahill Center for Astronomy and Astrophysics, California Institute of Technology, Pasadena, CA 91125, USA}
\email{}

\author{Obinna Modilim}
\affiliation{Center for Astrophysics $|$ Harvard \& Smithsonian, 60 Garden Street, Cambridge, MA 02138, USA}
\email{}

\author[0000-0002-6021-9421]{Nimesh Patel}
\affiliation{Center for Astrophysics $|$ Harvard \& Smithsonian, 60 Garden Street, Cambridge, MA 02138, USA}

\email{}

\author[0000-0002-2088-3125]{Kendrick Smith}
\affiliation{Perimeter Institute for Theoretical Physics, 31 Caroline St. North, Waterloo, ON N2L 2Y5, Canada}
\email{}

\author[0000-0001-6924-9072]{Lingzhen Zeng}
\affiliation{Center for Astrophysics $|$ Harvard \& Smithsonian, 60 Garden Street, Cambridge, MA 02138, USA}
\email{lingzhen.zeng@cfa.harvard.edu}

\begin{abstract}
Radio astronomy is uniquely coupled to exponential trends in 
computation because the optics (cross-correlation, beamforming, and imaging) and spectrometry 
(i.e. channelization) can now be done digitally.
Inexpensive analog-to-digital converters (ADCs) can sample signals from large numbers of antennas and graphics processing units (GPUs) allow us to coherently process 
wide-field radio data in real time, motivating large-$N$ aperture arrays at moderate cost.
We describe the 256-antenna Coherent All-Sky Monitor (CASM-256), 
a dense aperture array operating at 375-500\,MHz, currently being deployed at the Owens Valley Radio Observatory (OVRO) in Big Pine, California.
The large field-of-view (FoV$\sim10^4$\,deg$^2$) and point-source sensitivity of CASM-256 will allow it to detect local Universe fast radio bursts (FRBs). The nearby sample is 
ideal for unveiling the physical origin of FRBs, measuring the 
baryonic content of nearby galaxy halos, and discovering prompt multi-wavelength and multi-messenger counterparts to FRBs. CASM 
will search for fast transients in the Milky Way such as FRB analogs, pulsar giant pulses, and 
the new source class known as long-period radio transients.
We describe the instrument and 
present on-sky data from the first two dozen antennas, including an operational real-time GPU based FRB search pipeline. We 
emphasize the scalability of the concept and describe paths to a future CASM array with tens of thousands of antennas that could detect one million FRBs.

\end{abstract}



\section{Introduction} \label{sec:Intro}

Fast radio bursts (FRBs) are bright, impulsive transients 
whose physical origin is a major outstanding mystery in astrophysics \citep{lorimer2007, petroffreview, cordesreview}. 
Propagation effects imparted on the radio pulse contain cosmological information 
and are sensitive to astrophysical feedback. 
Fortunately, FRBs are abundant, appearing 
in a wide range of host environments from the 
local universe \citep{M81mohit,M81Kirst,ravi2025} to beyond redshift 2 \citep{caleb2025}. In the past decade, sensitive wide-field radio 
telescopes have been built that can discover and localize FRBs at increasing rates, enabled by low-cost, low-noise electronics and 
the exponential growth of compute hardware. The core technologies 
empowering MHz-GHz radio astronomy are driven by industry, adopted and optimized by the radio community for astrophysics experiments \citep{Hickish2016CASPER, weinreb}.

There are now 
many thousands of sub degree-scale localized FRBs \citep{Amiri2018CHIMEFRBSystem, CHIMEFRB2021Catalog1, CHIMEFRB2026Catalog2} and $\mathcal{O}(100)$ FRBs localized
with sufficient angular precision to obtain a host galaxy redshift. 
This sample has enabled a resolution to the decades-old ``missing baryon problem'' and constrain astrophysical feedback \citep{macquart2020, connor2025}.
These data have also led to new hints at plausible physical models
for the origin of FRBs. In short, FRBs appear to be a biased tracer of star formation \citep{sharma2025}, although some sources are found in older stellar environments \citep{M81mohit,M81Kirst,Eftekhari,Shah25}; $\lesssim10\%$ of FRBs are known to repeat and apparent non-repeaters are narrower, more broadband, and farther 
away than repeaters \citep{pleunis21}. A few repeaters are coincident with 
compact persistent radio sources \citep{MarcoteR1, niu2022, Moroianu}, though it is unclear if these are magnetar wind nebula, low-luminosity AGN, or something else entirely. At least one repeating source is periodic in its activity window, appearing in 
an ``on'' state for a few days every 16.3 days \citep{Li2020, pastor2020, pleunis180916}. The most likely origin of the coherent radio emission is in the 
magnetosphere of a neutron star \citep{kumar, nimmo2025}, 
although viable alternative models exist \citep{metzger2019, sridhar}.
Most of the key insights into the physical mystery of FRBs have come from 
non-cosmological sources at $<1$\,Gpc.

The detection of a MegaJansky burst from a known Galactic magnetar, 
SGR\,1935+2154, remains the decisive event connecting FRBs 
to known astrophysical sources. The burst was discovered 
in the far sidelobes of CHIME \citep{Andersen2020SGR1935CHIME}, and by the STARE2 telescope \citep{Bochenek2020STARE2}. The latter was designed to have a large FoV and 
only enough sensitivity to detect Galactic FRBs. 
FRB\,20200428 from SGR\,1935+2154 remains the only FRB-like source to have a multiwavelength transient counterpart \citep{Li2020HXMT}.
The Galactic event provided a strong motivation 
to build all-sky radio telescopes.
Based on this discovery, the Galactic Radio Explorer (GReX) \citep{connorgrex}
was built as a successor instrument to STARE2. GReX has antennas at OVRO, Hat Creek Observatory, Harvard, Cornell, and in Ireland \citep{shila2025}. With no new detections 
in two years of observing, GReX has placed further constraints on the rate of Galactic FRBs, showing that the true rate is below the modal Poissonian value implied by STARE2.

STARE2 and GReX are examples of incoherent all-sky monitors. 
In both cases, a single dual-polarization probe-fed circular waveguide antenna at 1.4\,GHz was used to search 
for Galactic radio bursts. Candidate information is combined 
across multiple stations incoherently to triangulate the pulse position after detection. With just one search beam, GReX 
can devote computing resources to searching at 
high time resolution ($\lesssim$50\,$\mu$s). However, the detection rate of a 
phased array of 
$N$ coherently combined antennas is $N^{1.5}$ times larger than a 
single radiometer, for Euclidean source counts. Additionally, the effective area of a typical antenna is proportional to $\lambda^2$, motivating lower frequencies. For example, the $8\,\sigma$ threshold of STARE2 
at 1.4\,GHz was 
10$^5$\,Jy. An array of 256 dual-pol dipoles at 450\,MHz has a detection 
threshold of 65\,Jy and a detection rate that is 
more than $5\times10^4$ times higher than STARE2 for extragalactic sources. 
Detecting FRBs outside of our Galaxy and measuring 
fainter Galactic analogs with an all-sky monitor requires an 
interferometric array. 

Most of the radio telescopes that have driven FRB discovery 
were not built for FRB science. 
CHIME was designed for 21\,cm cosmology \citep{Bandura2014CHIME}; 
Parkes, Arecibo, and Westerbork all precede FRBs by many decades 
\citep{Staveley-Parkes,Goldsmith2008Arecibo,Baars1973WSRT}; 
ASKAP and MeerKAT are multi-purpose survey instruments 
\citep{Johnston_2007,jonasmeerkat}. 
One exception to this trend was the DSA-110 at OVRO, 
whose large number of low-cost 4.65\,m dishes 
were purpose built for localizing FRBs 
\citep{koczdsa,ravidsa,lawdsa}.
The logical end-point of the
large-$N$, small-$D$ paradigm is the ``aperture array'', where individual 
antennas are pointed upwards without parabolic reflectors, maximizing instantaneous sky coverage. This is a natural design for a radio telescope that seeks to optimize fast transient science per dollar. 
While aperture arrays are as old as radio interferometry itself \citep{hewish}, 
only a small number of beams and narrow 
spectral bandwidths could be processed in the past. 
The Interplanetary Scintillation Array (built in 1967), 
which was designed by Antony Hewish and used by Jocelyn Bell 
to discover pulsars \citep{Hewish1968Pulsar}, had 2048 dipole antennas (the ``ISA-2000''?). Signals were beamformed by introducing cable delays to steer 
the telescope. On the Low-Frequency Array (LOFAR) in the Netherlands, stations of densely-packed antennas 
are beamformed in analog before digitization, limiting FoV 
to the per-station beam size \citep{mol2011lofarbeamformerimplementation}. The OVRO-LWA (12-85\,MHz) is able to image the entire visible sky instantaneously \citep{LWA}. In that case, its 352-antenna configuration is optimized for image quality and not fast-transient search, which requires a compact configuration to minimize the total number of beams.

An all-sky monitor that 
can coherently search its full primary beam of an individual dipole
($\sim10^4$\,deg$^2$) at millisecond timescales is only possible with modern compute back-ends and closely packed antenna configurations. In the limit of infinite compute, radio telescope design tends towards aperture arrays with 
enormous numbers of antennas, driving survey speed and all-sky capabilities.
To this end, several efforts are underway around the world 
to build all-sky arrays to search for FRBs and other short radio transients, such as BURSTT in Taiwan \citep{BURSTT} and its proposed expansion in Europe, 
CHARTS in Chile, and early concepts in Australia (CASATTA and CASPA \citep{CASPA}). 

We are building the Coherent All-Sky Monitor (CASM) at the 
Owens Valley Radio Observatory (OVRO). A pathfinder array of 
256 dual-polarization printed circuit board (PCB) dipoles is 
currently being deployed on the north-south arm of a tee-shaped 
railroad track at OVRO\footnote{https://casm-telescope.com}. The CASM concept is unusually scalable. 
Unlike optical telescopes whose cost scales steeply with aperture ($\propto D^{2.5}$ for ground-based and $\propto D^{3.5}$ for space-based 
observatories \citep{van_Belle_2004})
the cost scaling of CASM is sub-linear in the number of antennas.
A more sensitive CASM simply requires more GPUs and 
more digitizers, 
both of which grow in capability and decrease in cost with time.

\begin{figure}[htbp]
    \centering
    \includegraphics[width=\textwidth]{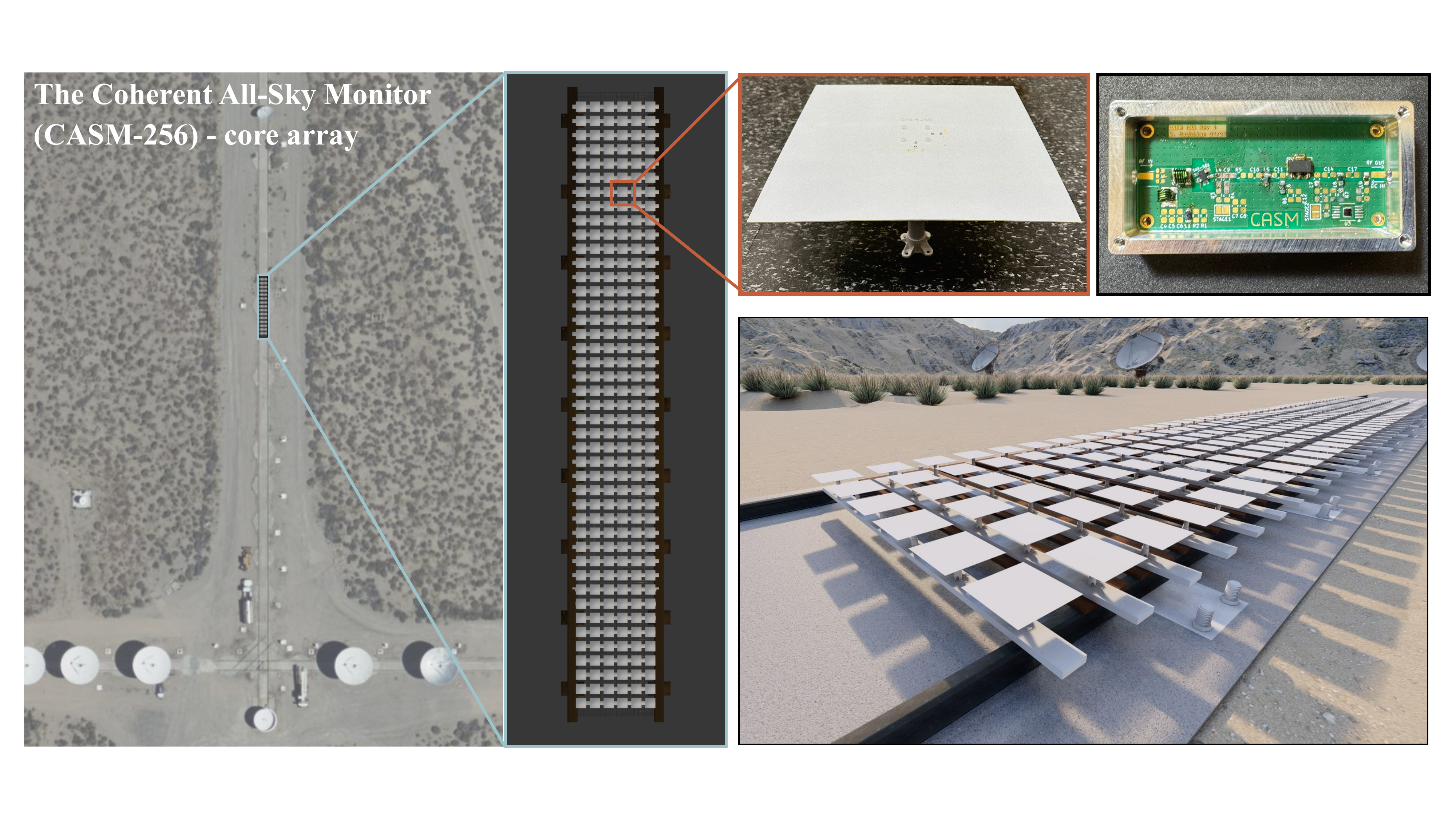}
    \caption{\textbf{The 256-antenna Coherent All-Sky Monitor (CASM-256)
      under construction at the Owens Valley Radio Observatory (OVRO).} A 20$\times$3\,m core of 256-antennas at 375--500\,MHz will search for fast transients and localize with outrigger stations (not shown, currently under design). North is up on left panel satellite image. The bottom right is a rendering of 
      the core array without radomes. The top right panels show the PCB antenna and 
      its low-noise amplifier, both custom built for CASM.}
    \label{fig:main}
\end{figure}

\section{Scientific motivation}
We focus here on primary science cases for the 256-antenna CASM that is currently under deployment. A more sensitive CASM 
with $\gtrsim10^4$ antennas would have a much broader scientific 
domain, particularly if the array configuration is jointly optimized for imaging. 

\subsection{A local Universe sample to study the origin of FRBs} 
    FRBs are 
    brief, bright, and ubiquitous, but most events to date have come from beyond $\sim$\,1\,Gpc ($z \gtrsim0.3$) where it is difficult to study their local environment. As with other transient classes, 
    nearby FRBs have been disproportionally informative in 
    understanding their origin.
    In addition to the Galactic FRB\,20200428, a repeating source was pinpointed to a globular cluster
    in M81 (just $\sim$3.6\,Mpc away) \citep{M81mohit, M81Kirst}. This fact is difficult to reconcile with 
    FRBs arising from the recent death of massive stars. At the median redshift of localized FRBs, it would have been impossible to confirm a globular cluster origin. 
    The periodically active 
    repeating FRB\,20180916B is offset from a 
    star forming region in a spiral galaxy at 150\,Mpc \citep{MarcoteR3}. 
    Recently, \citet{blanchard} obtained deep NIR follow-up 
    imaging of FRB\,20250316A with JWST, showing that the source is 
    on the outskirts of an HII region and coincident with a 
    point source candidate at $m_{F150W2} \approx 31.0$\,mag and $m_{F322W2} \approx 30.4$\,mag.
    The host galaxy, NGC\,4141, is just 40\,Mpc away. CASM-256 
    will localize a nearby sample of FRBs that are bright and rare---only detectable with a large FoV due to the fleeting nature of FRBs. 
    Many Northern repeating FRBs will spend $\sim10$ hours per day in our primary beams, allowing for 
    dense sampling of repetition statistics and 
    periodicity searches that are not possible with 
    existing facilities. 
    
\subsection{Measuring the baryon content of the circumgalactic medium}
The bulk properties of the circumgalactic medium (CGM) are poorly understood, despite the central role 
it plays in galaxy evolution, feedback, and the mediation of gas between 
the IGM and galaxies \citep{Tumlinson_2017}. FRBs provide an opportunity to measure $\int n_e\,dl$ for $L*$ galaxy halos, thereby constraining the total mass of the CGM directly for the first time.
No other probe (X-ray, UV spectroscopy, t/kSZ, etc.) can measure the column of baryons independently of metallicity, temperature, or velocity \citep{prochaska18}.
The intergalactic medium (IGM) dominates the DM of most FRBs discovered to date \citep{connor2025}, with $\mathrm{DM_{IGM}}\approx 860\,z$. Sources within $100$\,Mpc ($z<0.025$) will receive an 
average DM of $\lesssim$\,25\,pc\,cm$^{-3}$ from the IGM. 
At such distances, intersection of a foreground galaxy halo or cluster is very unlikely. This means the total DM of local Universe FRBs is dominated by 
the Milky Way and the host galaxy contribution,

\begin{equation}
    \rm DM_{obs} \approx DM_{MW,ISM} + DM_{MW, CGM} + DM_{h, ISM} + DM_{h, CGM}.
\end{equation}

\noindent The Galactic ISM component is well modeled by Milky Way pulsar 
DMs, particularly off of the plane where model uncertainty is 
$\sim$\,10\,pc\,cm$^{-3}$ \citep{ocker2020}. Therefore, 
a sample of nearby FRBs will allow us to jointly constrain the CGM 
of the Milky Way and of host galaxies in a range of halo masses, assuming 
$\rm DM_{h, ISM}$ can be modeled. This has already proven constraining 
with existing nearby FRBs \citep{cookfrb2023, ravi2025, leungCGM}. The total DM of FRB\,20220319D at 50\,Mpc 
was very low, allowing \citet{ravi2025} to place an upper limit on the 
gas content of the Milky Way halo that was significantly below 
the cosmological average of $\frac{\Omega_b}{\Omega_m}$. \citet{leungCGM} 
recently showed that the inferred $\rm DM_h$ of $z<0.20$ FRBs are inconsistent with hydrodynamical simulations in which feedback is weak. Finally, 
\citet{mccarty2026} showed that the $10^{13}$\,M$_\odot$ elliptical host galaxy contributes no detectable DM despite a prediction of $>200$\,pc\,cm$^{-3}$ in 
no-feedback scenarios. 
These results hint 
at strong feedback in which galaxies do not retain all of 
their baryons, consistent with recent results from kSZ, tSZ, weak lensing, and X-rays \citep{boryana, bigwood, seigel25} as well as cosmological FRBs \citep{connor2025, reischke, sharma32}.
CASM-256 will produce a sample of $\mathcal{O}(50)$ FRBs within 200\,Mpc, 
hosted by galaxies with different stellar masses and at a range of 
inclination angles. This will allow us to model $\rm DM_{h, ISM}$ and 
``weigh'' the CGM directly.

\begin{figure}[htbp]
    \centering
    \includegraphics[width=0.75\textwidth]{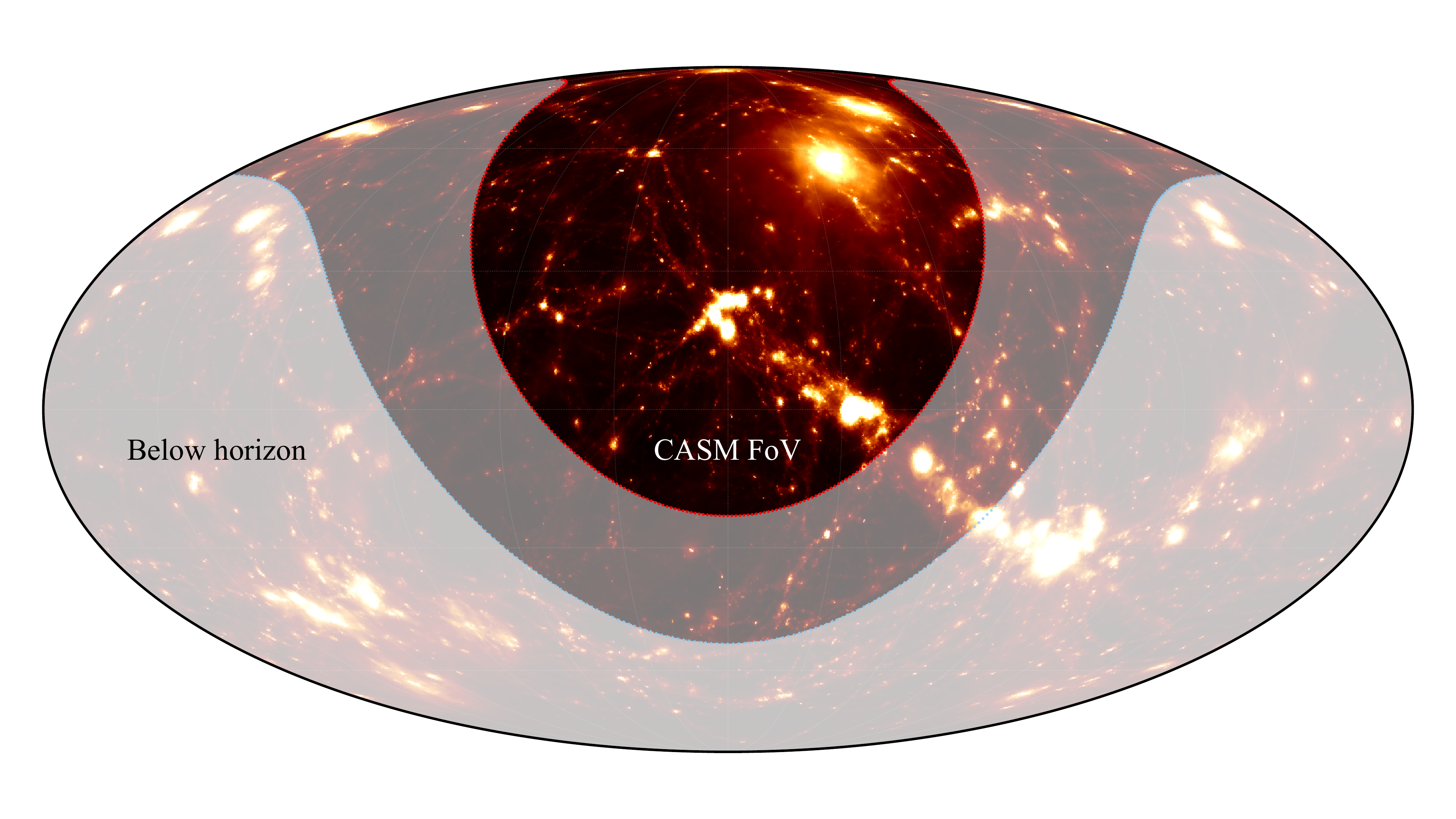}
    \caption{\textbf{The CASM field-of-view.} CASM-256 will continuously search the Northern sky for fast transients between 375 and 500\,MHz over $\sim$7,500\,deg$^2$. The background image is the cosmic baryon column at $z\leq0.01$ from ray tracing in the Illustris TNG-300 simulation \citep{konietzka2025}. The DMs of CASM-256 FRBs will measure the diffuse gas in halos with little IGM contribution.}
    \label{fig:fov}
\end{figure}

\subsection{Coincident multi-wavelength \& multi-messenger emission}
No transient multi-wavelength emission has been detected from an extragalactic FRB. Upper-limits have been placed in the optical, 
infrared, and X-ray for repeating FRBs, but the majority of 
bursts to date have not had multi-wavelength co-observing \citep{nicastro, Hiramatsu}. Unsurprisingly, the 
one example of FRB-like emission coincident with a high energy transient was 
SGR\,1935+2154, by far the closest source at just $\sim$\,7\,kpc away. 
That X-ray transient would not have been detectable outside of our galaxy. An all-sky 
telescope like CASM-256 offers two distinct advantages over 
more sensitive arrays with smaller FoV. CASM-256 will observe 
$\sim$\,10$^4$\,deg$^2$ continuously, meaning we will overlap with transient 
surveys at shorter wavelengths enabling coincident constraints. It will also detect bright, nearby events 
that have a better chance of reaching the detection threshold of 
all-sky X-ray monitors and fast optical cameras. A co-detection of a high energy 
fast transient with an extragalactic FRB is a ``holy grail'' for understanding 
the emission mechanism \citep{Bransgrove26}.

Theorists have predicted prompt or precursor radio emission from 
merging neutron star binaries and black hole/neutron star binaries since before the first LIGO detection \citep{hansen, Pshirkov, Totani_2013}. 
Multi-messenger FRBs would establish a new channel for coherent extragalactic radio emission, because coalescing compact objects cannot account for more than a few percent of the volumetric rate of FRBs \citep{connor2016a, ravirepeat, shin23}. An offline search for FRB GW associations was carried out by LIGO using a subset of CHIME/FRB sources, placing upper-limits on the event rate \citep{LIGOScientific:2022jpr}. 
During a three-year run with O5 sensitivities, LVK could see hundreds of GW events involving at least one neutron star \citep{lvkkunnum}. Half of Northern Hemisphere events will be in our beam. 

Roughly 500\,meters from CASM is another all-sky telescope: The OVRO Long Wavelength Array (LWA-352) \citep{LWA}. The LWA operates at 12-85 MHz and is optimized 
for high quality continuous imaging of the whole visible sky at 10\,s cadence. It also has the ability to trigger on real-time transients,
with a detection threshold of 100 Jy ms (7$\sigma$) fluence within the 55–85 MHz band. \citep{nikita2025}. 
An FRB with DM=250\,pc\,cm$^{-3}$ will arrive at the top of
LWA's band roughly 2.5 minutes after detection by CASM. We will trigger LWA with CASM FRBs. Both will trigger on external multi-wavelength and multi-messenger events. We expect that 
some nearby FRBs will be detected by both CASM and OVRO-LWA. The periodically 
active repeating source FRB\,20180916B was measured down to 100\,MHz with 
fluences spanning 40--300\,Jy\,ms \citep{pastor2020, pleunis180916}, above the LWA threshold assuming the emission continues down to 80\,MHz. 

\subsection{Strong gravitational lensing}
A small fraction ($\sim10^{-3}$) of FRBs will be strongly lensed by 
intervening dark matter \citep{Mu_oz_2016, leung-2022, Connor_2023}. FRBs are unique as 
time-domain lensing sources because they are short-lived, coherent, and cosmological. This allows 
for temporal interferometry that can identify lensing time delays below even their millisecond burst widths \citep{Eichler2017, Wucknitz2021, kader-2022}. For FRB lensing events where the deflector is $>10^{8}$\,M$_\odot$ ($t_{grav} \gtrsim 0.5$\,hr), radio telescopes like CHIME and DSA cannot detect the lensed copy because they must be pointing 
at the exact position when the second copy arrives. 
Deflector mass modeling is not precise enough to anticipate its arrival to within a pointing or transit. Therefore, 
an all-sky monitor like CASM is the only way to achieve 
a significant probability of finding FRBs strongly lensed 
by massive halos. While the redshift distribution 
of CASM-256 is too shallow for appreciable lensing 
optical depth, an array with thousands of elements will 
find strong lenses with sub-millisecond precision 
on the lensing time delays up to years \citep{Connor_2023}.

\subsection{Fast Galactic transients}
SGR\,1935+2154 has emitted coherent radio emission 
over a large range of energies, from 100\,mJy to 
1.5\,MJy \citep{Kirsten2021SGR1935,Andersen2020SGR1935CHIME,Bochenek2020STARE2}. If other Galactic magnetars have a wide luminosity 
function extending above 100\,Jy, CASM-256 will fill in the 
parameter space between normal radio magnetars and FRBs. It 
will also discover rare, super-giant pulses from radio pulsars 
thanks to the long semi-continuous exposure. 

Long period radio transients (LPTs) have recently 
been discovered with periodicity between 18\,min and 6.45\,hr with 
pulse durations of 0.37--300\,s \citep{Hurley_Walker_2022, rea2026longperiodtransientslpts, Lee_2025}. Their 
exact origin is an open question, but two of the 
$\sim10$ known objects are confirmed white dwarf binaries \citep{de_Ruiter_2025, rodriguez}. 
CASM-256 will be sensitive to pulses above 2.3\,Jy for a 
1\,s burst and 0.3\,Jy for a 1\,min burst. Northern LPTs 
will spend upwards of 10 hours in our primary beam each 
day, allowing us to discover new sources and monitor 
known objects with dense temporal sampling.

\subsection{Synergies with facilities at other wavelengths}
Argus is a 1,200-telescope optical array that will cover $\sim$\,8,000 deg$^2$ per exposure \citep{Law_2022, corbett2022skyterabitsecondarchitecture}, nearly identical 
instantaneous sky coverage to CASM. Argus is fully funded, with plans to be on sky within several years. If the Argus Array is built in the Southwestern US, half of FRBs detected by CASM will have simultaneous optical data from Argus, without needing to coordinate pointing or trigger. Conversely, fast optical transients will produce either new upper-limits or the first radio/optical detections. 
Argus reaches $g\sim19.6$ per minute and has a sub-second fast mode. This is an entirely new 
capability in FRB science. Previously, FRBs could only 
be followed up in optical if they were repeating sources. 
Additionally, CASM is designed to discover nearby FRBs
for which Argus will have a better chance of detecting 
coincident emission. 

Stellar radio transients represent another promising synergy between CASM and Argus. Violent stellar eruptions on active and young stars can produce coherent swept-frequency radio bursts analogous to solar Type II and Type III bursts \citep{Vedantham2020}. At CASM frequencies, these bursts probe particle acceleration and coronal mass ejections on other stars. These processes are well-studied on the Sun but poorly constrained elsewhere. Simultaneous Argus optical light curves would identify the associated white-light flares, enabling the first large statistical study of the radio/optical flare connection beyond the Sun \citep{Davis2025}. Long-period radio transients with stellar counterparts could also be cross-matched with Argus photometry to constrain their progenitor systems. Because both instruments monitor the same sky continuously, this science does not require physical coordination.

Together with OVRO-LWA, these three all-sky instruments will span wavelengths from 6\,m to 700 nm, providing the most complete multi-wavelength view of fast transients to date. All three telescopes will also overlap with LIGO/Virgo/KAGRA O5, enabling simultaneous
radio, optical, and gravitational-wave coverage of neutron star merger events.

\subsection{Technology pathfinder}
Compelling science can be done with the 256-element array.
In addition to its scientific motivation, 
CASM will be a technology pathfinder. As compute capabilities grow exponentially, aperture arrays with 
enormous numbers of antennas will become more attractive. 
The exact CASM-256 design 
would not scale to tens or hundreds of thousands of antennas cost effectively, 
but its commissioning will allow us to 
optimize for a large future array with low per-channel cost. Its software pipeline is scalable, particularly 
CASM's use of FFT beamforming, low-precision GPU-based 
single pulse search, and high-speed networking.

    \begin{table}[h!]
    \centering
    \renewcommand{\arraystretch}{1} 
    \setlength{\tabcolsep}{12pt}      
    \rowcolors{2}{gray!10}{white}     
    \begin{tabular}{l l}
    \toprule
    \textbf{Specification} & \textbf{Value} \\ 
    \midrule
    Location               & Owens Valley Radio Observatory (OVRO) \\ 
    Number of antennas     & 256 (core) + outriggers \\ 
    Frequency Range        & 375--500 MHz (93\,MHz processed)\\ 
    Field of View (FoV)    & 7,500 deg² \\
    8\,$\sigma$ detection threshold (1\,ms / 1\,s)    & 65\,Jy / 2\,Jy \\ 
    Digital Back-End       & FPGA-based F-engine, GPU-Based X-Engine\\ 
    Search beams           & 1020 stationary beams \\ 
    Time resolution        & 1\,ms \\ 
    Estimated Detection Rate & 0.5--2 FRBs/week \\ 
    Localization Precision & $\sim$\,5-7" (OVRO) $\leq$\,0.1" (VLBI) \\
    \bottomrule
    \end{tabular}
    \caption{Specifications for CASM-256}
    \end{table}

\section{Array design}

The CASM-256 core is a semi-regular 43$\times$6 grid of 
antennas built on the north-south arm of the 
site's railroad track, built previously for configurable interferometers at OVRO. Each of the 43 rows of antennas consists of a 3\,m vinyl fence plank with 6 antennas that sit inside of three plastic tote radomes. The north-most antenna plank will only have 4 antennas, for a total of 256. In the north-south direction, the 43 antennas are separated by 0.5\,m. 
In the east-west direction, the spacing of the 6 antennas 
is [0.45\,m, 0.38\,m, 0.45\,m, 0.38\,m, 0.45\,m]. The separations are 
set by physical constraints within the antenna radomes.
The array is therefore a rectangle with 
dimensions 21.5\,m$\times$2.1\,m. The unusual antenna layout meets two key criteria for the array. First, they are closely packed to limit the number of beams required to tile the FoV. The required number of beams is,

\begin{equation}
    N_{beam}\approx \left (\frac{4\pi}{\mathrm{FoV}}\frac{D_{EW}D_{NS}}{\lambda^2}\right),
\end{equation}

\noindent where $D_{EW}$ and $D_{NS}$ are the east-west and north-south dimensions of the array, respectively, and FoV is the 
primary beam size in radians. For CASM-256, this number is 
roughly 500 beams. If the array were 1\,km by 1\,km, we would need 
to form and search 10$^7$ beams, requiring over 10,000 times 
more search compute with our current algorithm. The next criterion was that we did not want to 
disturb ground, pour concrete, or build significant new infrastructure, in order to minimize build time. This is why we 
took advantage of the existing railroad track that 
was level and raised above the ground.

The core array of 256 antennas has effectively no ability to 
localize astrometrically, beyond large nearby galaxies like M31. Its synthesized beams are $\sim$\,2\,deg$\times$20\,deg. To localize FRBs 
to their host galaxy, and eventually to their exact position 
within their galaxy, we will build ``outrigger'' stations. 
Outriggers can be triggered to save phase-preserving voltage data
after an FRB is detected in the core array, enabling offline 
interferometric localization.
This will be done in two stages, starting with on-site 
outriggers each with 6 antennas. The maximum baseline of 2\,km is dictated by the perimeter of OVRO, providing $5''$ astrometry for $S/N\geq10$ and sufficient precision to identify a host galaxy if it is closer than $\sim$\,300\,Mpc. Next, we plan to build semi-autonomous
VLBI outrigger stations of 30--60 antennas. 

\begin{figure}[htbp]
    \centering
    \includegraphics[width=\textwidth]{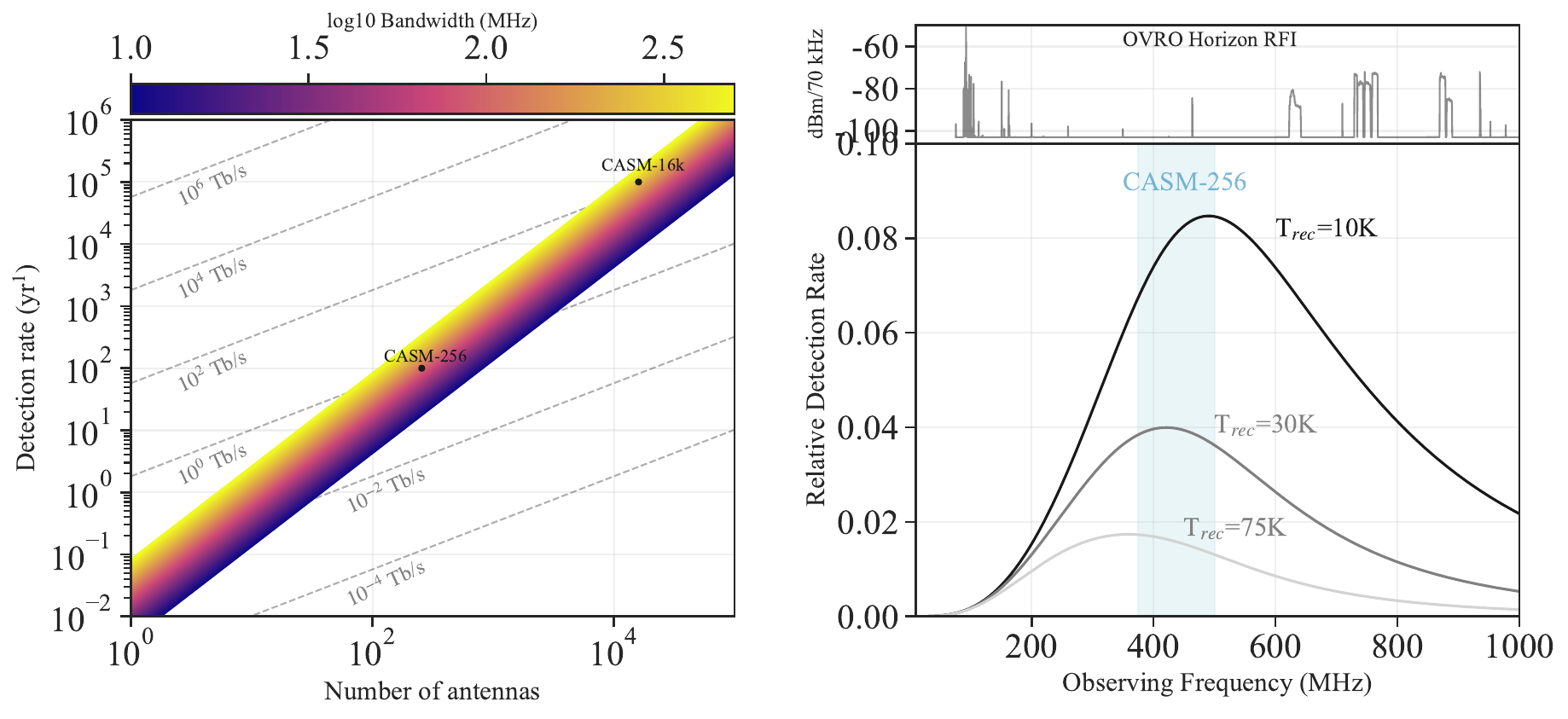}
    \caption{The expected CASM detection rate as a function of 
    number of antennas (left) and observing frequency (right). The left panel shows annual FRB detections colored by 
    radio bandwidth from 10\,MHz (purple) to 500\,MHz (yellow). 
    Lines of constant data rate are plotted, where 
    we assume dual-polarization feeds, 8\,bit data, 
    and Nyquist sampling (i.e. rate is 32\,$N_{ant}\,B$ in bits/s). Points 
    on those curves that are not in the colored region 
    have unrealistically low or high bandwidth. The right 
    panel shows relative detection rate vs. frequency for 
    three different receiver temperatures along with the 
    horizon RFI spectrum at OVRO.}
    \label{fig:rate}
\end{figure}

\subsection{Sensitivity and expected rate}

We first estimate the sensitivity of a 256-element array operating between 400--500\,MHz with full-width half maximum (FWHM) of 100\,deg and a FoV of $\Omega\approx7500$\,deg$^2$. The System Equivalent Flux Density (SEFD) of such an interferometer is 
given by the ratio of system temperature to gain, $T_{\mathrm{sys}}/G$. This
can be thought of as the noise RMS in Jy for 1 second of integration time and 1\,Hz of bandwidth. 
The gain of the array is 
$N_{ant}\,G_i$, where each dipole gain is $G_i\sim\frac{4\pi}{\Omega} = 7.5$\,dBi for the aforementioned FoV, assuming 100\% efficiency. The effective area is given by,

\begin{equation}
    A_e = \frac{N_{ant}\,G_i\,\lambda^2}{4\pi}.
\end{equation}

\noindent At 67\,cm wavelength, the effective area of CASM-256 is
51\,m$^2$. This is slightly smaller than a single ASKAP dish, but with $\sim$\,300$\times$ larger FoV. Assuming an LNA temperature of 20\,K and a mean sky temperature of 20\,K across the band \citep{deOliveiraCosta2008}, the SEFD is 
2750\,Jy and the noise RMS is $\sigma_s = \frac{\mathrm{SEFD}}{\sqrt{n_p\Delta\nu\,\tau}}\approx6.8$\,Jy for $\tau=1$\,ms and 90\,MHz of usable bandwidth. Therefore, the array's 8\,$\sigma$ detection limit would be a fluence of roughly 65\,Jy\,ms for 1\,ms bursts. 

We scale from CHIME/FRB using Euclidean source counts to estimate the total detection rate of CASM \citep{connor2016b}, assuming the two telescopes have roughly equivalent $T_{sys}$.
Assuming CHIME has a $\sim$200\,deg$^2$ FoV, a sky-averaged SEFD of 75\,Jy over 300\,MHz of usable 
bandwidth, and a detection rate of 1,000 FRBs per year, 

\begin{equation}
    \mathcal{R_{\mathrm{CASM}}} \approx 65 \, \left ( \frac{N_{ant}}{256} \right )^{3/2}
    \left ( \frac{B}{100\, \rm MHz} \right )^{3/4}\,\mathrm{yr^{-1}}.
\end{equation}

\noindent We expect the 256-antenna pathfinder array to 
discover 0.5 to 2 local Universe FRBs per week. The factor of 4 in rate uncertainty includes effects like beamforming coherence, final system temperature, and RFI. We plan to quantify these effects once the full CASM-256 array is deployed.

An array with 
32,000 dipoles would find $10^6$ FRBs in $\sim$\,5 years on sky. Even with 100$\times$ more effective collecting area, ``CASM-32k'' would still be in the Euclidean scaling regime.
In the left panel of Figure~\ref{fig:rate} we show the 
annual detection rate as a function of number of dual-pol antennas. 
The colored lines show the detection rates for different radio bandwidths, 
spanning 10\,MHz to 500\,MHz. Interestingly, for a fixed 
data rate, one should maximize the number of antennas and minimize 
bandwidth because $\mathcal{R}\propto N_{ant}^{1.5}$ whereas 
$\mathcal{R}\propto B^{0.75}$. For FRBs, one still needs enough 
coverage in $\Delta\lambda^2$ to detect the dispersed pulse, meaning 5,000 antennas and 1\,MHz of total bandwidth is not optimal unless pure image-plane searches are deployed. The point is further complicated by the non-linear cost of 
correlation and beamforming in $N_{ant}$, assuming 
a non-FFT (i.e. brute force) approach to those pipeline stages. Algorithms like beamforming and correlation are now limited by memory bandwidth and total memory on GPUs, meaning FLOPS are not the ultimate bottleneck in 
many cases. Caveats aside, 
we opt for just 100\,MHz on CASM-256 partly for the advantage of 
$N_{ant}$ over $B$.

For a Euclidean volume, the mean source distance scales 
as the square root of sensitivity, due to the inverse square law \citep{dongzi_2018}, even if the shape of the observed distance distribution, $\frac{\mathrm{d}n}{dz}$, can look very different depending on the FRB luminosity function and detection systematics \citep{jamespdmz}. With $\sim$\,50 times less point-source sensitivity than CHIME, we expect the mean FRB redshift from the 256-antenna CASM to be $\sqrt{50}$ times lower than the CHIME sample's mean redshift. Most CHIME/FRB sources do not have a host galaxy redshift, but we can apply the DM/redshift relation from \citet{connor2025} to the observed DM$_{ex}$ distribution in CHIME. We find the mean DM-inferred redshift of CHIME FRBs is $\left < z_{\rm DM} \right > \approx 0.20$. This value agrees with the 
mean redshift of localized FRBs from the DSA-110 ($\left < z \right > = 0.28$) 
and ASKAP ($\left < z \right > \approx 0.25$). 
Thus, the mean redshift in the CASM sample would be $\sim$\,0.03 or a 
comoving radial distance of about 130\,Mpc, based on this scaling.

\subsection{Choice of frequency band}
As discussed, the effective collecting area of a dipole is proportional to $\lambda^2$, 
driving the array towards low frequencies.
This is balanced by sky temperature increasing strongly towards 
long wavelengths, where telescopes are sky noise dominated below $\sim$\,400\,MHz 
due to the frequency dependence of Galactic synchrotron ($T_{sky}\propto\nu^{-2.5}$).
Low frequencies have other challenges such as the deleterious effects of 
scattering, intrachannel dispersion smearing, and long dispersion delays 
\citep{connor_2019}. In choosing the CASM band, we attempted to optimize these considerations along with the local RFI environment of OVRO. Below, we step through these considerations.

Point-source sensitivity is given by Gain, system temperature, 
and usable bandwidth. Gain is determined by the antenna design and observing 
frequency. The system temperature is,

\begin{equation}
    T_{sys} = T_{rec} + T_{sky}(\nu).
\end{equation}

\noindent Here, the receiver temperature is dominated by the LNA noise temperature with 
contributions from any impedance mismatch in the analog chain. 
A low $T_{rec}$ pushes the preferred observing frequency upwards, where the 
low $T_{sky}$ is advantageous; large $T_{rec}$ pushes the band downwards, 
because the $\nu^{-2}$ collecting area term becomes relatively more important (Figure~\ref{fig:rate}).
The signal-to-noise recovery after temporal smearing is $S/N_{obs} = \eta_{smear}\, S/N$, where,

\begin{equation}
    \eta_{smear} = \frac{t_{FRB}}{\sqrt{t^2_{FRB} + \tau^2 + 
    t^2_{s} + t^2_{DM}}}.
\end{equation}

\noindent Here, $t_{FRB}$ is the timescale of the FRB, 
$\tau$ is scattering time, proportional to $\nu^{-4}$, 
$t_{DM}$ is intrachannel dispersion smearing, which scales as $\nu^{-3}$, and $t_s$ is sampling time. Loss of signal increases towards low frequencies, and the detection rate becomes $\propto \left (\eta(\nu)\,\frac{G}{T_{sys}}\sqrt{B} \right)^{1.5}$ \citep{connor_2019}.

In the right panel of Figure~\ref{fig:rate}, we show the detection rate 
of CASM-256 for different receiver temperatures. We have assumed a 
typical FRB is 1\,ms in duration with a DM of 500\,pc\,cm$^{-3}$, and a scattering 
timescale of 100\,$\mu$s at 600\,MHz. We assume 30\,kHz frequency channels over 100\,MHz of bandwidth, and a flat intrinsic FRB rate. 
We also plot the horizon RFI environment at OVRO between 0-1\,GHz. There is a clean window between 375-500\,MHz. 
We have chosen to sample this band, using 93\,MHz of 
bandwidth as our science band. This will likely be 
390-483\,MHz.

\section{Hardware}

\subsection{Array structure}
We have built a 25$\times$3\,m wood frame within the OVRO railroad track 
to support a metal ground screen, on top of which the antennas sit. The CASM ground screen is made from three 100\,ft rolls of 
hardware cloth that was pulled taut and stapled into the wood 
frame. 43 planks supporting 6 antennas each are to be placed on the array frame and ground screen. This modularity allows us to remove and service antenna planks without walking on the array ground screen. These planks are 10\,ft vinyl fence posts made from PVC. They will be separated by roughly 0.5\,m in the north-south direction. The hollow fence post acts as conduit for the 12 coaxial cables (two per polarization), 
which are routed from the LNA at the base of our antennas through a
hole in the plank, terminating at the west edge of the core array.
LMR195 coaxial cables then take the signal from the edge of the array to an underground vault that houses our analog back-end boards and the F-engine hardware. 

\subsection{Antenna}

\begin{figure}[!ht]
    \centering
    \includegraphics[width=0.45\textwidth]{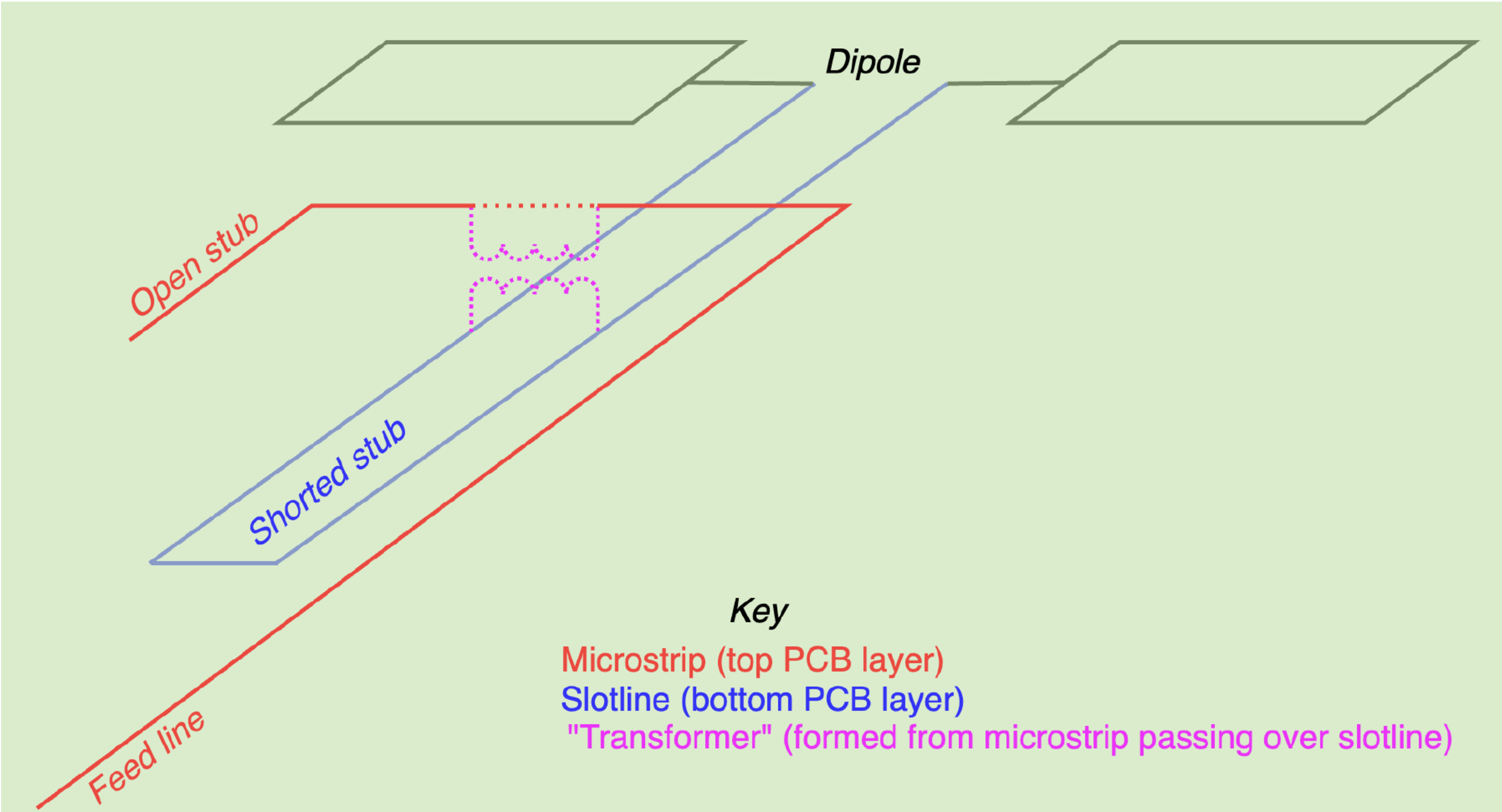}
    \caption{The schematic circuit for the CASM printed dipole feed network \citep{Gutierrez2025}. The signal propagates along the microstrip feed line and is coupled to the dipole antenna. The open stub at the end of the microstrip and the shorted stub in the dipole antenna are optimized to achieve the best coupling efficiency.}
    \label{fig:ant_circuit}
\end{figure}

\begin{figure}[htbp]
    \centering
    \includegraphics[width=0.45\textwidth]{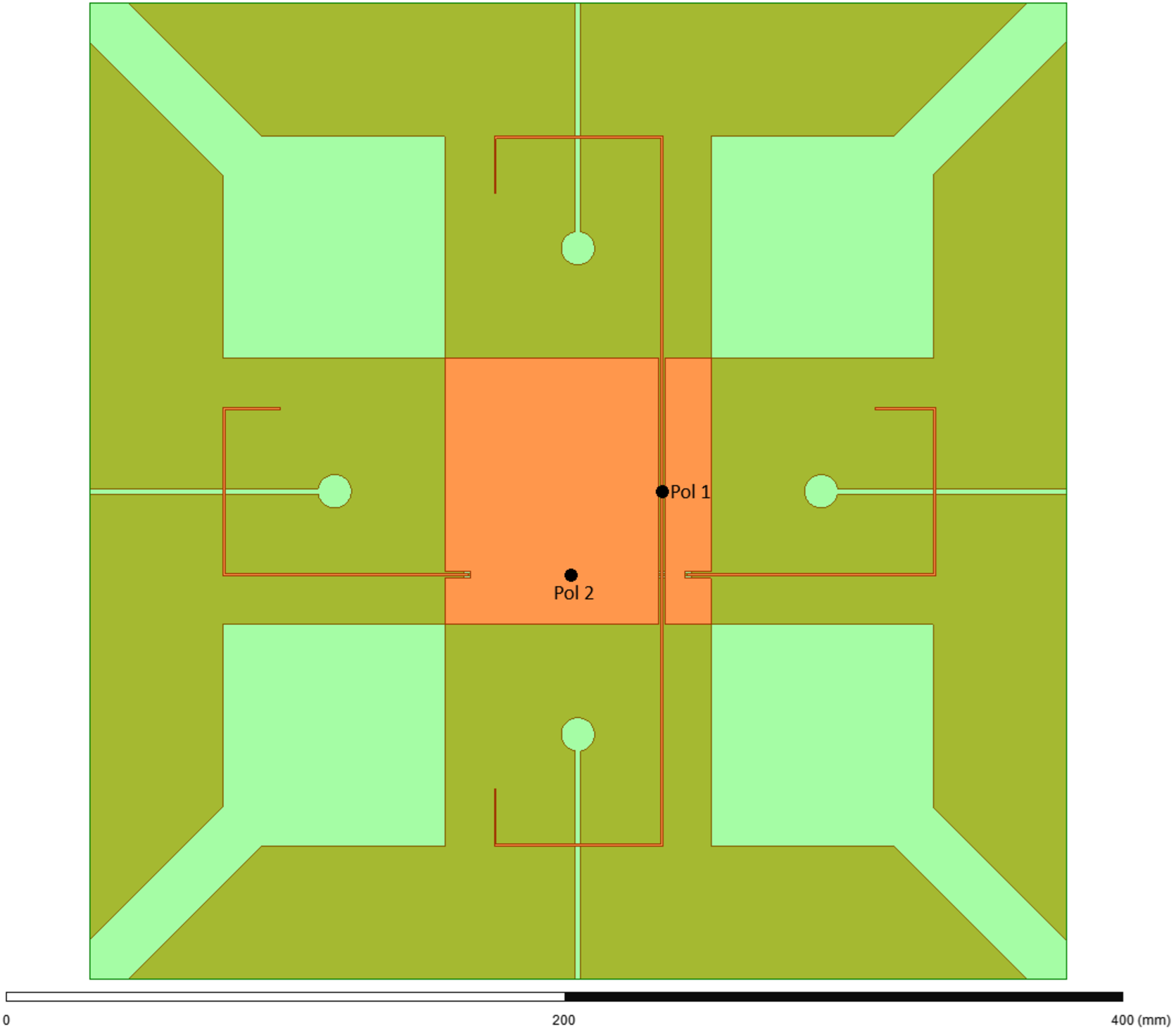}
    \caption{The CASM antenna model consists of four sets of printed dipoles and feed lines integrated into a planar PCB, forming a dual-polarization antenna. The antenna is excited through the SMA feed points labeled as `Pol 1' and `Pol 2' in the figure.}
    \label{fig:ant_model}
\end{figure}

\begin{figure}[htbp]
    \centering
    \includegraphics[width=0.85\textwidth]{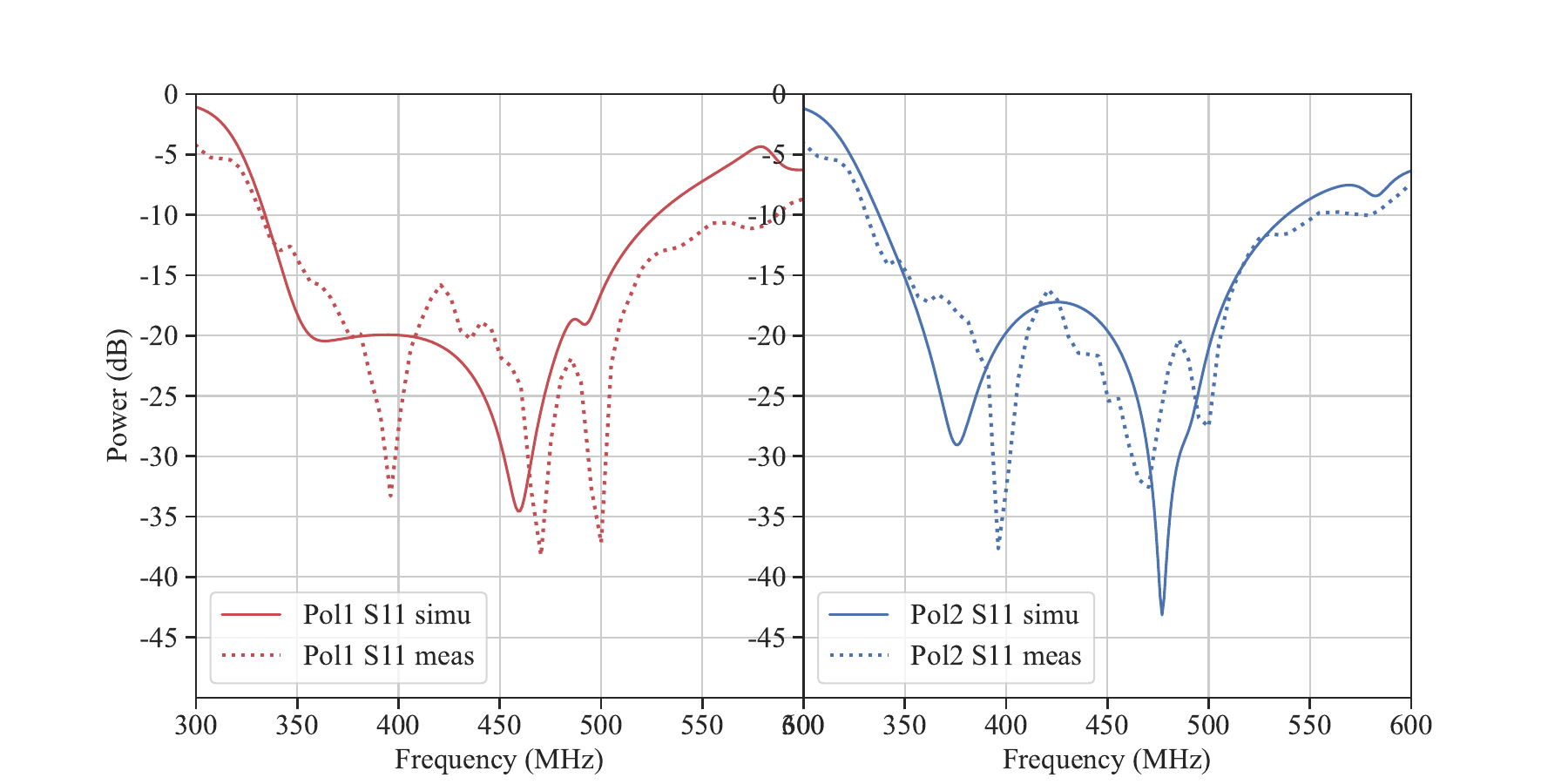}
    \caption{Simulated and measured return loss performance for both polarizations is shown. From 350 MHz to 515 MHz, the return loss is better than -15 dB. The -10 dB bandwidth covers the frequency from 330 MHz to 570 MHz, yielding a fractional bandwidth greater than 50\%.}
    \label{fig:ant_rl}
\end{figure}

The CASM antenna is based on the designs proposed in \cite{8477070} and \cite{4907127}. It features a dual-polarization printed dipole design printed on a PCB. Each polarization consists of two pairs of dipoles fed by parallel microstrip lines  \citep{Gutierrez2025}. Figure~\ref{fig:ant_circuit} illustrates the schematic circuit for a printed dipole. The antenna is optimized by tuning the open stub and shorted stub to achieve best coupling between the feed line and the dipoles. Figure~\ref{fig:ant_model} shows the CASM antenna model. The antenna is printed on an FR-4–based PCB that is 1.6 mm thick and 35 cm on each side, approximately one quarter of a wavelength. The feed points have a standard impedance of 50 $\Omega$, and the corresponding feed lines are designed for 100 $\Omega$. The simulated and measured return loss for both polarizations is shown in Figure~\ref{fig:ant_rl}. The antenna achieves a –15 dB bandwidth from 350 MHz to 515 MHz (about 38\% fractional bandwidth) and a –10 dB bandwidth from 330 MHz to 570 MHz (greater than 50\% fractional bandwidth). 

\subsection{Analog electronics}

The CASM low noise amplifier (LNA) provides 35\,dB gain at 12\,K noise temperature over the 375-500\,MHz frequency band using a two-stage, unconditionally stable design. The Mini Circuits SAV-581+ PHEMT transistor is used in the first stage for its extraordinarily low noise figure, accompanied by a low-loss, lumped input match. This notoriously unstable transistor is stabilized with distributed inductance on the source and resistive loading on the drain, yielding unconditional stability at both its input and output. The Mini Circuits PGA-105+ MMIC amplifier is used for the second stage, significantly simplifying the LNA layout and avoiding complex stability and power matching. This output feeds directly into a bias tee, which enables DC biasing the LNA over coaxial cable from backend analog electronics. One of these custom 
LNAs is shown in Figure~\ref{fig:main}.

\section{Digital backend}

\subsection{F-engine}
\begin{figure}[htbp]
    \centering
    \includegraphics[width=0.65\textwidth]{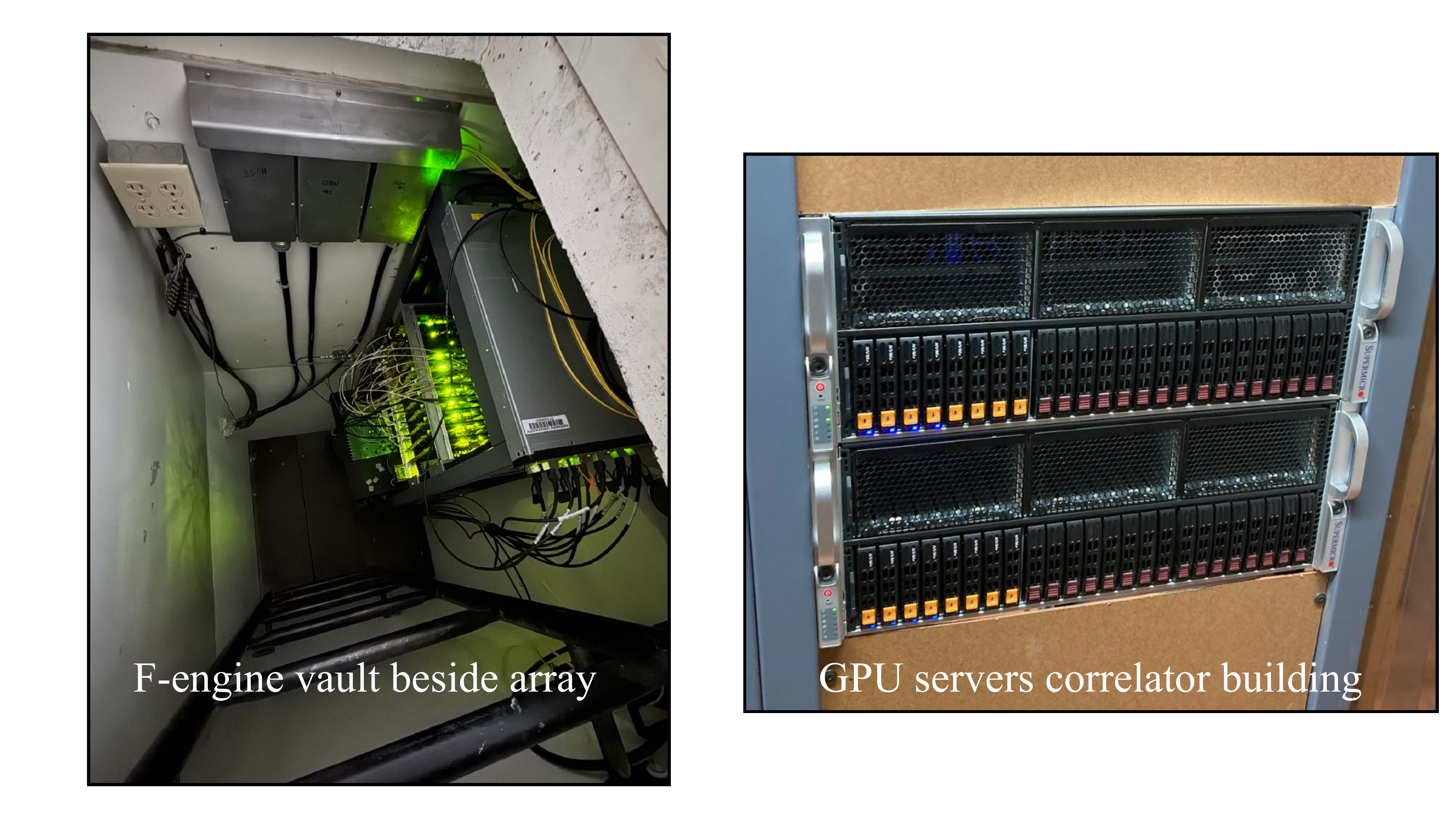}
    \caption{\textbf{Compute hardware for CASM-256}. The left panel shows an underground vault beside the core array. It houses 43 SNAP boards, a 100\,Gbe switch, fiber connection to the control building, and a clock distribution system. The right panel shows two GPU servers, each with two dual-port 200\,Gbe NICs and 10 RTX\,4000 Ada GPUs. These are in a control building roughly 300\,m from the core array.}
    \label{fig:fengine}
\end{figure}

The CASM F-engine is responsible for digitizing our signal, 
channelization, and streaming voltage data to our GPU servers 
via a network switch and fiber link. The F-engine hardware resides in an existing underground vault beside the railroad tracks that 
already has fiber connections to the correlator building, power, and RF shielding (see Figure~\ref{fig:fengine}). 

Signals from the backboard are digitized and channelized 
on Smart Network ADC Processor (SNAP) boards. A custom 
FPGA image was built by Real-Time Radio Systems Ltd\footnote{realtimeradio.co.uk} to use all 12 ADC inputs on the boards. 
The 256$\times$2 antenna/polarization signals are spread out 
over 43 SNAP boards, with one plank of 6 antennas arriving 
at one SNAP.
Each input is sampled at 250\,MS/s with a 4096-channel PFB (Hamming, 4-tap), producing 
4+4bit real/imaginary voltage data over spectral channels covering 125\,MHz. Pre-ADC analog filtering selects the 
fourth Nyquist zone (375-500\,MHz) for direct sampling.
We send 93\,MHz of the 125\,MHz sampled bandwidth 
over 10\,Gb links to a Mellanox Spectrum SN3420 networking switch, 
where data are routed onto 6$\times$100\,Gb fiber links. These fibers 
connect the vault beside the array to the correlator room, roughly 300\,m away. 

\subsection{Clock \& synchronization}

Each SNAP board in the vault is referenced to a GPS-disciplined oscillator (GPSDO) that provides both a 10\,MHz frequency reference and a 1 PPS timing pulse. PPS and the 10\,MHz signal is carried from the correlator building to the CASM vault over fiber. The 10 MHz connects to one Valon 5009B Synthesizer in the vault, which generates a 250\,MHz sampling clock for the ADC and FPGA, allowing digitization of 125\,MHz of bandwidth. The Valon's output is amplified and split to 43 SNAPs. The 1 PPS signal is routed to a digital input on the SNAPs and used to arm the FPGA when a common PPS rising edge arrives. 

\begin{figure}[htbp]
    \centering
    \includegraphics[width=0.65\textwidth]{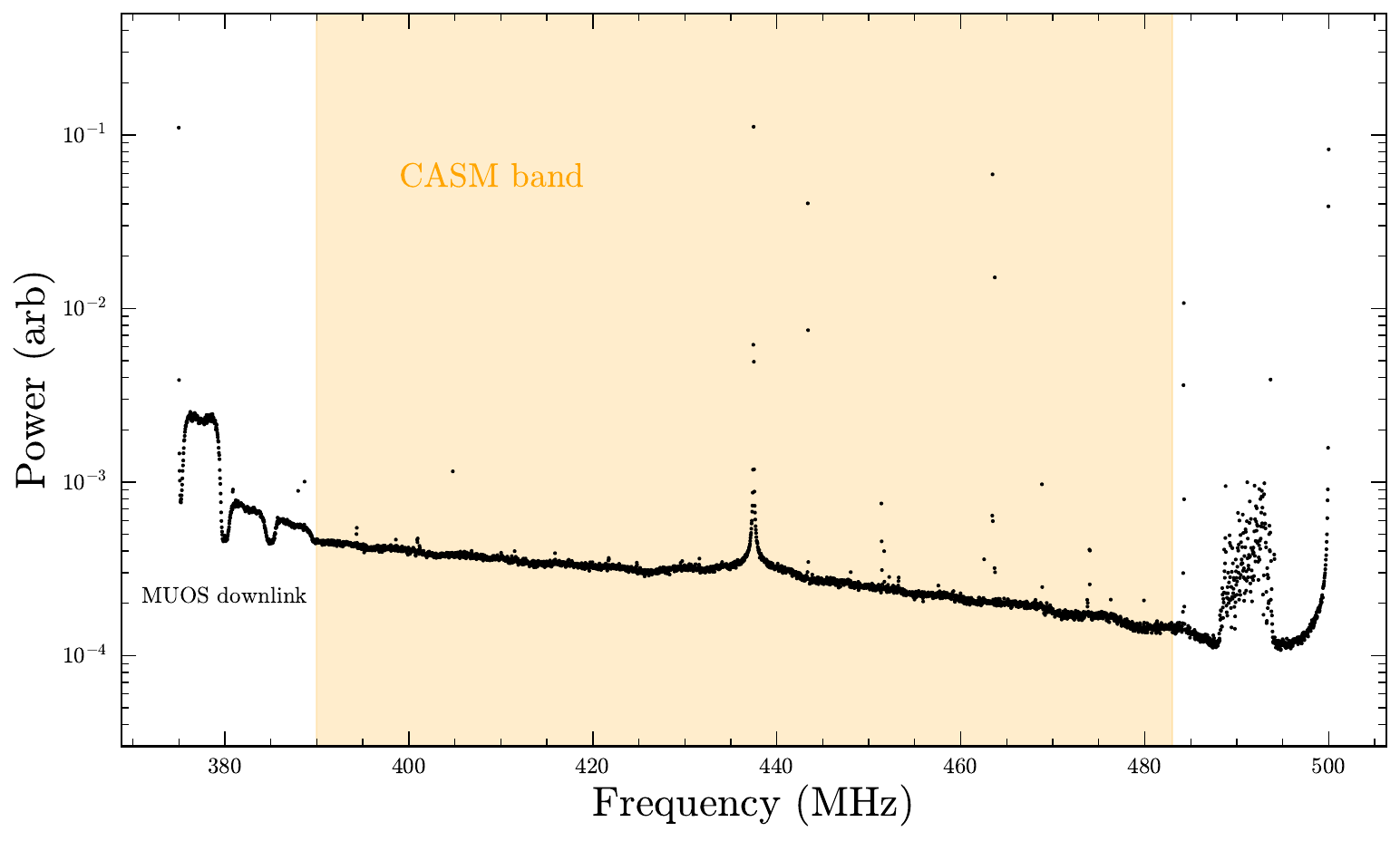}
    \caption{\textbf{The measured digital spectrum from one CASM-256 antenna.} Our F-engine produces a spectrum every 32\,$\mu$s with 4096 channels from 375-500\,MHz. We send 3072 contiguous channels off the SNAP board within that range, limited by the 10\,Gbe SFP+ port on the board, but single-ADC spectra can be obtained for the full range. The plotted spectrum is a typical 1.6\,s integration. Our chosen 93\,MHz band is shown in the orange shaded region. Haystack RFI at the bottom of the band is the MUOS satellite downlink (360–380\,MHz, mirrored by aliasing). The peak at 437.5\,MHz is a digital spur caused by a voltage offset between the ADCs on our SNAP boards.}
    \label{fig:spectrum}
\end{figure}

\subsection{GPU servers}

The GPU servers are responsible for packet capture, 
beamforming, cross-correlation, imaging, and single-pulse search. 
The whole data processing pipeline for CASM-256  
takes place on two compute nodes, each with 
10 RTX\,4000 Ada GPUs. Each server has 
2\,TB of memory and 4$\times$7.6 TB disks for storage. There are two dual-port NDR200 Mellanox CX7 NICs (QSFP-112), meaning each server can in principle consume 800\,Gb/s for a total of 1.6\,Tb/s. The total data rate 
arriving at the servers from the vault is 387\,Gb/s, which is spread out over 6 NICs in total (i.e. $\sim$\,65\,Gb/s per 200\,GbE NIC). 

\section{FRB localization in two stages}

Localization is paramount to all  
FRB science. For many FRB applications (cosmic baryons, strong lensing, EMGW co-detection of LIGO events, etc.), one must achieve physical localization precision at the 10\,kpc level to identify 
a host galaxy and obtain its redshift. In recent years, it has become clear that studies of the physical origin of FRBs benefit from even better physical localization, often requiring 
Very Long Baseline Interferometry (VLBI) \citep{marcote2022preciselocalizationsrepeatingfast,frbcollaboration2025chimefrboutriggersdesignoverview}. We plan to build localization outriggers of CASM-256 in two stages: An on-site test outrigger array that can achieve $\sim$\,5$''$ 
astrometric precision. And later, low-cost, semi-autonomous VLBI 
outrigger stations for $\lesssim0.1''$ precision. In both cases, we will search for FRBs using the CASM core array of 256 antennas, triggering the buffered raw voltage data at our outriggers if an FRB, Galactic event, or other signal of interest is detected. 

\subsection{On-site outriggers} 

Outrigger stations tracing the perimeter of OVRO were built for the DSA-110.  
These were fifteen 5\,m dishes to complement the $\sim$\,90 antennas 
that make up the DSA-110 core array.
The project required trenching and laying hundreds of kilometers of 
optical fiber that could carry analog signal to the correlator 
room for offline interferometric localization. 
On CASM-256, we hope to take advantage of this infrastructure by placing 
CASM antennas near DSA-110 outriggers. 
Each antenna station has excess power and spare fiber that CASM could use. 
A vinyl plank of 6 dual-polarization antennas could be placed roughly 
10\,m from the DSA-110 outrigger dishes. Each would be connected 
to a single 12-input SNAP board in an RF- and weather-proofed enclosure.
That signal would be digitized and sent over fiber by 
a 10\,Gb/10\,km optical transceiver.
    
The longest on-site baseline will be roughly 2.5\,km, allowing for 5-7\,'' localization precision of a 
10\,$\sigma$ FRB. This would not be sufficient to identify a host galaxy at $z=0.5$. Fortunately, CASM-256 will find nearby FRBs. At 300\,Mpc, 5'' corresponds to less than 10\,kpc. The outrigger data can also be streamed to the correlator to produce real-time all-sky images, in search of radio emission from stars, 
planets, and Galactic long-period radio transients. 

The design and installation effort for on-site outriggers is less significant than our planned 
off-site VLBI stations. For this reason, we plan to first test our localization capabilities at OVRO. An RF-proof box will be used 
to house a SNAP board, Valon 5009B, DC power supply, and 
12 back-end boards. Critical to this effort will 
be establishing an RFI proof system that does not impact the CASM outrigger antennas nor other telescopes at the observatory.  

\subsection{Off-site VLBI outriggers}

Non-repeating FRBs have been routinely localized to their host galaxy with 
arcsecond precision for the past
$\sim$\,five years \citep{ravidsa, bannister2019}. It has proven extremely valuable to localize FRBs with sub-arcsecond 
precision \citep{MarcoteR1, MarcoteR3}, where the immediate environment can be studied in detail. The PRECISE 
collaboration has led this profitable effort \citep{marcote2022preciselocalizationsrepeatingfast} and CHIME/VLBI is bringing it to scale \citep{leungvlbi,frbcollaboration2025chimefrboutriggersdesignoverview, tone}.
This was exemplified by recent JWST observations of a VLBI-localized FRB at 
40\,Mpc in NGC\,4141 \citep{blanchard}, which was a very rare event: It was the highest S/N CHIME event in seven years of observing, and one of the most nearby bursts ever discovered. CASM-256 will provide dozens of such sources (mean $z\approx0.03$), thanks to its large FoV and modest sensitivity. 

VLBI outrigger stations that consume less than $1.5$\,kilowatts, do not require large fiber networks, and can share data over a Starlink internet connection could be placed at two to four sites across North America to achieve sub-arcsecond localization. Each station ought to have at least $10\%$ of the sensitivity of the 
256-antenna core array. We therefore aim to have 5 planks of 6 antennas at each site, requiring 5 SNAP boards, totaling 100 Watts of power for the per-station F-engine. The compute can be quite slim because we will not be beamforming, correlating, nor searching the data locally. We simply need packet capture that can handle the $\leq50$\,Gb/s data rate, streaming to disks with enough capacity to handle one hour's worth of data. Once an FRB is detected and confirmed, the locally-stored voltage data would be sent via Starlink to 
the OVRO correlator where it will be localized offline.

A compact network switch will have at least 5 SFP+ connections from the SNAP boards and output a single 100 Gb QSFP uplink. The server need only a 100\,Gbe QSFP NIC that can capture the 50\,Gb/s and 
disks that can be written to quickly after a trigger from the CASM core array. It would need enough RAM to buffer 45 seconds of data, meaning $\sim$\,512\,GB. The machine would be a short-depth 1U 
field server made of custom components. 
Assuming only power is available at the 
VLBI outrigger site, Starlink will provide internet connection for controlling the station and sending 
data after a detection. If an FRB is detected and triggered 
once every 48 hours, the uplink data rate must be high enough 
to send both FRB data and calibration data. The FRB data is 
modest, because we can ``cut out'' the dispersed pulse, which amounts to just 0.5\,GB if we use $\pm50$\,ms around the burst. Calibration data, however, requires enough sensitivity to detect standard VLBI calibrators. Based on the experience of CHIME/VLBI, we expect to send 
30 seconds of data near the time of the event. This would be about 
1 hour of transfer time every several days, assuming 50\,MB/s on Starlink (Business / Priority).

A key motivation in designing light-weight semi-autonomous 
outrigger station is flexibility in site choice. We will explore traditional observatory options, such as Hat Creek in Northern California and NRAO in Socorro, New Mexico. 
Finally, we will explore the option of placing outriggers at the homes of willing participants. The VLBI outrigger design remains under consideration.

\section{Software pipeline}

The CASM-256 software pipeline captures complex voltage data 
arriving from the vault at 387\,Gb/s, beamforms those data to create $\mathcal{O}(10^3)$ beams, searches the beams for dispersed single pulse candidates, and then clusters/filters/classifies the FRB candidates resulting in a real-time trigger of our outrigger antennas. The pipeline must also continuously form visibilities with which we can generate beamformer weights and all-sky images at $\sim$\,seconds cadence. Below we describe each major component of the pipeline. Several core components of the real-time pipeline have been built by Fourier Space\footnote{https://fourierspace.com.au/about/} and deployed for testing by the CASM collaboration. The full 512-input 
system with 1024 beams, DM$_{max}=1000$\,pc\,cm$^{-3}$, and a real-time correlator has been shown to run on our two-server backend.

\subsection{Packet capture} \label{subsec:packet_capture}

The digitized signals from the SNAP boards are transmitted to the GPU servers as a set of six User Datagram Protocol (UDP) data streams, with each stream containing data from 512 of the 3072 frequency channels.
Each stream is switched to one of the NICs on the two GPU servers, with each server receiving three of the six subbands.
The packets are transferred from the NICs to a staging buffer in RAM via direct memory access (DMA), and the packet capture software assembles the data from these staged packets into blocks of 2048 time samples, which are elements of a large ring buffer holding approximately $35\,\mathrm{s}$ of voltage data.
These 2048-sample blocks are then consumed by the beamformer/correlator pipeline, and are also periodically consumed by a separate application to produce high-level monitoring plots for every antenna.

\begin{figure}[htbp]
    \centering
    \includegraphics[width=1\textwidth]{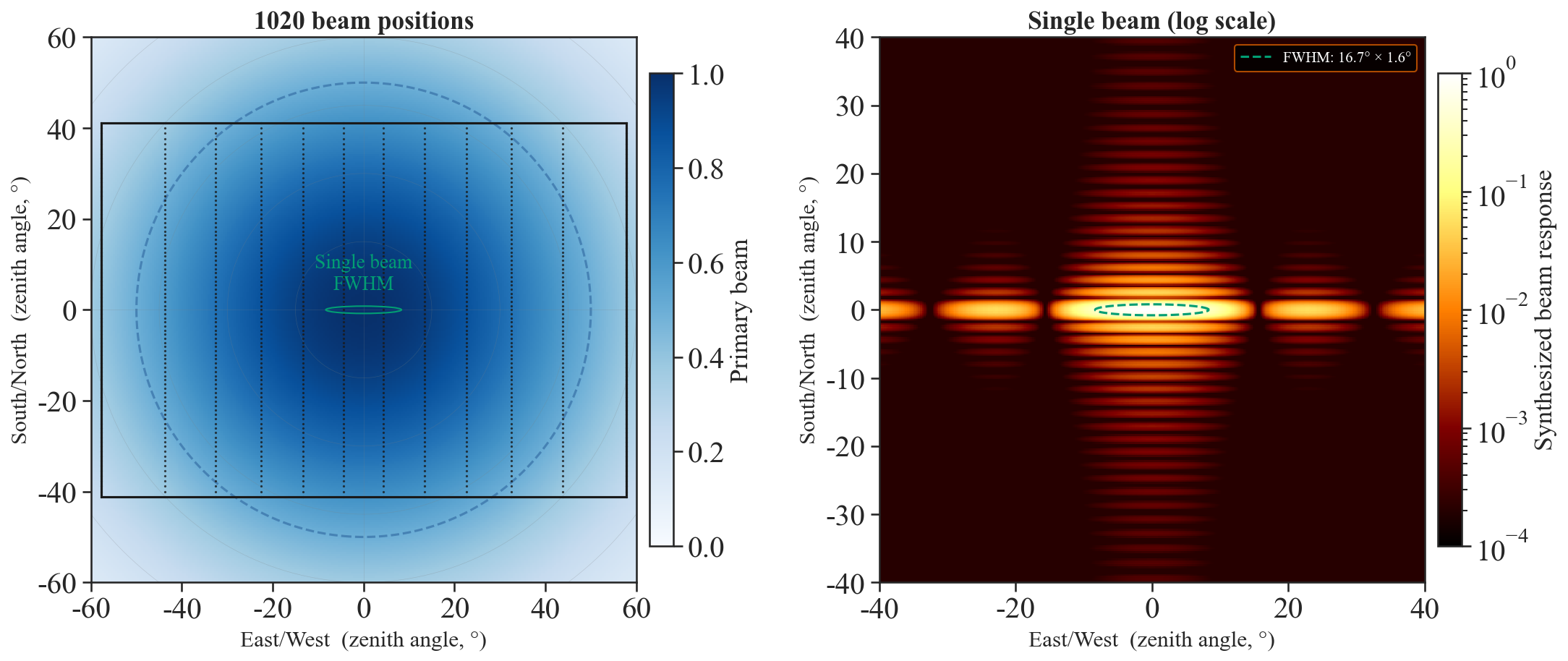}
    \caption{\textbf{Example sky positions of 1020 beams from the FFT beamformer (black dots, left) and a single synthesized beam for the CASM-256 layout (right)}. The FFT beamformer oversamples such that arbitrary beams can be formed from linear combinations of the 3072 intensity beams. In this case, we show 1020 beams placed on a grid, but other configurations are possible. This figure uses the 450\,MHz PSF.}
    \label{fig:beams}
\end{figure}

\subsection{Beamforming}

Beamforming requires summing complex voltage data (proportional to the incident electric field) across antennas with weights chosen to maximize gain 
in a particular direction on the sky. For an all-sky telescope like CASM, 
we wish to form a beam at all independent positions in the primary beam. ``Brute-force'' matrix based beamforming requires roughly $N_{b}\,N_{chan}\,N_{ant}\,t_{samp}^{-1}$ operations per second, 
where $N_b$ is number of beams, $N_{chan}$ is the number of frequency channels, and $t_{samp}$ is the sampling time of channelized voltage data. A dense array requires roughly $N_{ant}$ beams to cover its FoV, resulting in the familiar scaling that the brute-force beamforming cost is $\propto N_{ant}^2$. However, if antennas lie on a regular 
grid, beamforming and cross-correlation can be achieved via an FFT \citep{peterson2004forecastepochofreionizationviewableprimeval, tegmark}. 
FFT beamforming effects a significant reduction in FLOPS, scaling as $\mathcal{O}(N_{ant}\log N_{ant})$ rather than $\mathcal{O}(N_{ant}^2)$. In practice, such algorithms are often memory bandwidth limited due to their low arithmetic intensity. 
If memory bandwidth issues can be mitigated, FFT beamforming is a major
opportunity to save on compute for regular arrays like CASM. It makes beamforming on ultra-large arrays tractable by requiring $\frac{N_{ant}}{\log N_{ant}}$ times fewer FLOPS (more than 1000x savings for $N_{ant}>10^4$). BURSTT carries out multi-stage beamforming, with a first stage on Xilinx RFSoC FPGA boards and a second-stage on CPUs \citep{LinBURSTT}. Our beamforming system is purely GPU-based.

A custom FFT beamforming pipeline was built by Kendrick Smith 
for the CASM-256 array\footnote{https://github.com/Coherent-All-Sky-Monitor/casm\_bf}. 
CASM’s long axis is uniformly spaced with antennas separated by 0.5\,m, enabling us to FFT beamform in the north-south direction. The east-west axis is length-6 and non-uniform, with spacings (0.45\,m, 0.38\,m, 0.45\,m, 0.38\,m). The length-43 axes are padded to 128 points, which is FFT'd to produce a (128, 6) complex array of 
voltage fan beams. The algorithm then uses dense matrix based beamforming 
in the east-west direction to produce the full (128,24) grid of voltage beams 
that tile the entire CASM FoV. The electric field/voltage array is then squared, summed over two polarizations, and downsampled in time to 1\,ms. This oversampled set of 3072 ``basis'' beams can then be combined linearly to 
point at arbitrary positions with 4-by-4 bicubic interpolation. 
The algorithm has been implemented in a highly-optimized megakernel that takes in time/frequency voltage data and outputs beamformed intensity timestreams for all frequency channels. The kernel
minimizes GPU memory transfers and avoids writing intermediate data products (e.g. post-FFT partial beams or full basis beams) to global GPU memory, alleviating GPU memory bandwidth considerations. 

The interpolation step of the beamforming algorithm approximates ``exact'' beamforming. We have shown that the correlation coefficients between this approximation and the exact solution never falls below 0.997, which is more than sufficient 
for fast transient science. Captured data are placed in global memory on 6 of 20 GPUs. Each GPU processes all antennas and 512 of 3072 frequency channels.
The per-GPU load fraction of CASM's CUDA beamforming implementation is roughly 40$\%$.  
This is for $32\,\mu$s input data, 512 channels, and 1024 beams with NVIDIA's RTX 4000 Ada GPUs. In other words, the FFT implementation effectively 
fits the whole CASM-256 beamformer on just 2.5 workstation GPUs. Significant 
improvements could be made for a $>$\,$10^4$ antenna array by switching to float16 
and tensor core FFTs, in addition to speedup from improved GPU hardware. 
Beamforming a CASM-32k would likely fit on fewer than 50 GPUs.

\begin{figure}[htbp]
    \centering
    \includegraphics[width=\textwidth]{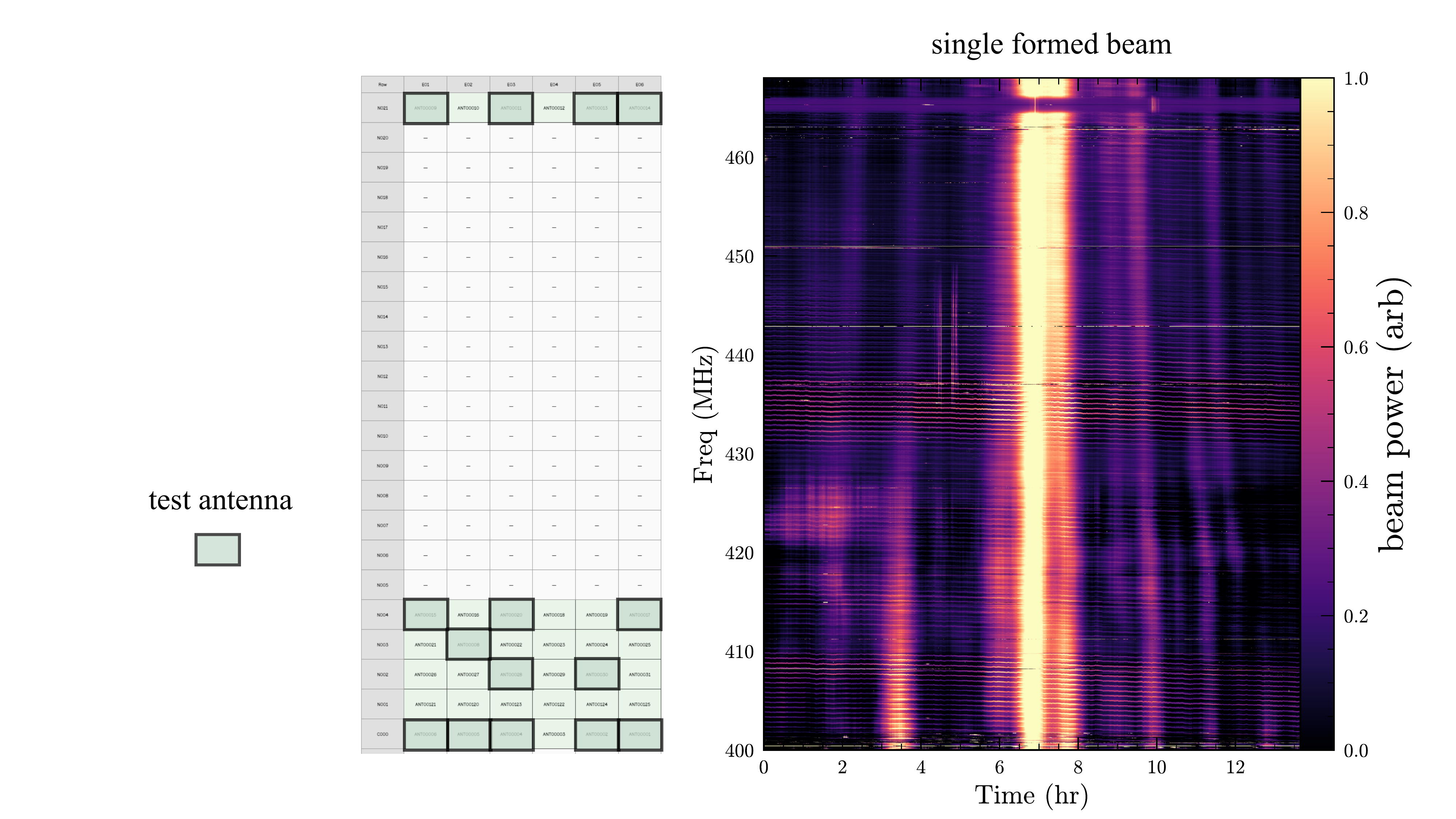}
    \caption{\textbf{Output from the real-time beamformer during a solar transit, written with $t_{samp}=4$s}. The left panel shows the array configuration for commissioning data in this work. This is the top half of the 256-antenna array. Filled boxes are antennas that have been deployed, filled boxes with black edges have operational analog chains. Chromatic sidelobes can be seen in the right panel.}
    \label{fig:beamformer}
\end{figure}

Our custom FFT beamformer trades flexibility for speed. We have also written a direct (non-FFT) voltage-based beamformer 
that operates on an arbitrary number of antenna inputs and 
antenna positions. We are currently using this beamformer 
for commissioning purposes because the FFT beamformer expects a grid of 43$\times$6 antennas. The direct beamformer uses CUDA Templates for Linear Algebra Subroutines and Solvers (CUTLASS) tools for 
its matrix multiplication. While not as fast as the FFT beamformer, our CUTLASS beamformer is still highly performant, operating at ${\sim}100\,\mathrm{TOPS}$ --- the matrix operations are performed using the available tensor cores on RTX 4000 Ada GPUs, with the input voltages and weights represented as 8-bit integers and the output beamformed voltages stored as 32-bit integers.
The output from both the FFT beamformer and the direct beamformer is converted to a 16-bit floating point representation before being saved to host memory for corner turning and processing by the single-pulse search pipeline (Section \ref{subsec:single_pulse}).
In Figure~\ref{fig:beamformer}, we show data from the real-time direct beamformer. 

\begin{figure}[htbp]
    \centering
    \includegraphics[width=0.9\textwidth]{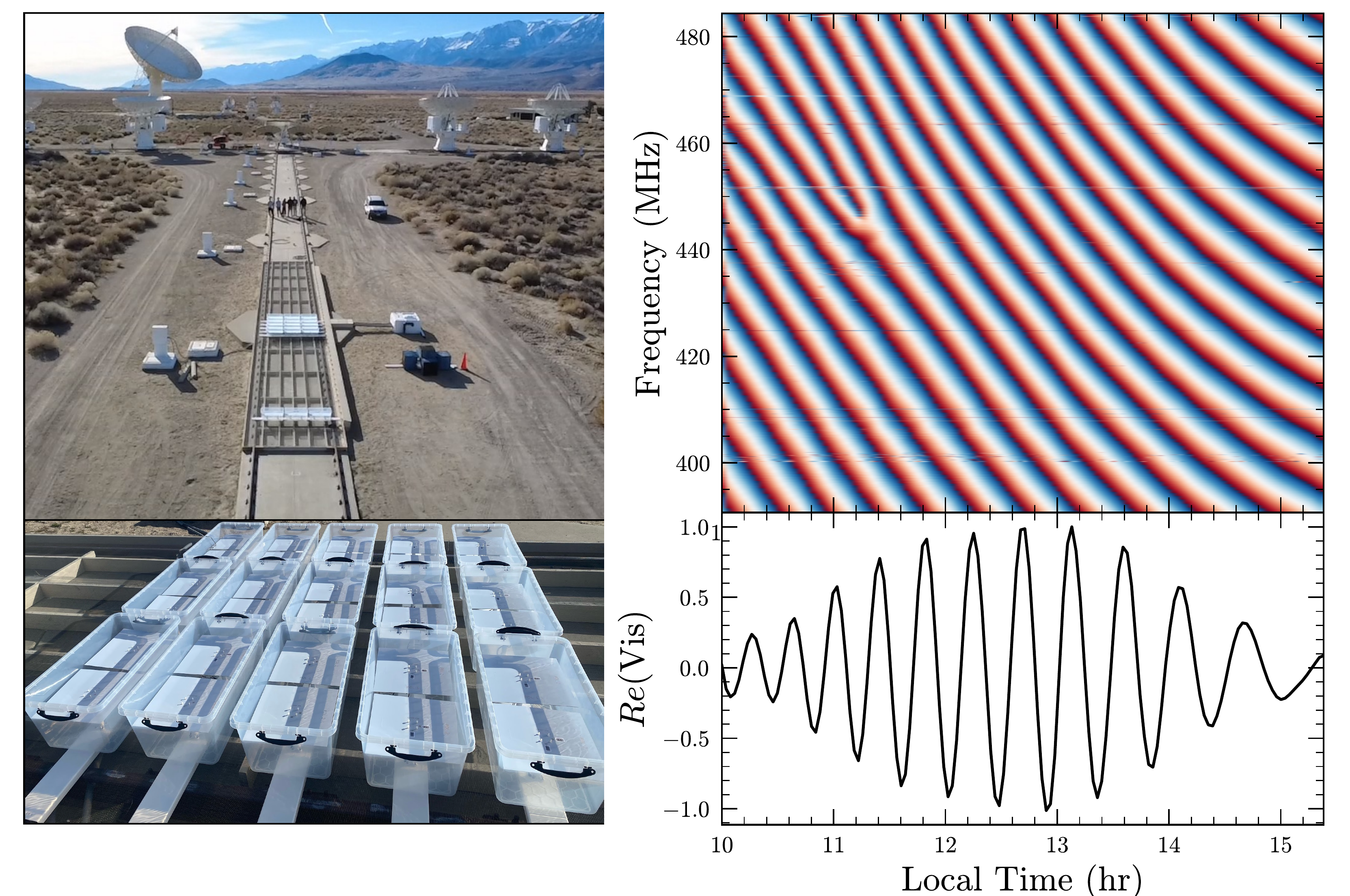}
    \caption{\textbf{The first core antennas and fringes from CASM-256 at OVRO.} Images on the left show the groundscreen
    built atop the north-south railroad and the first 37 antennas (six planks plus one antenna beside the core for testing). 
    The right figures show fringes of the Sun for a 
    single baseline. The top right panel is visibility phase vs. time and frequency; the bottom right panel is the real component of visibility as a function of time during solar transit, in a 30.5\,kHz channel at 444\,MHz.}
    \label{fig:array}
\end{figure}

\subsection{Cross-correlation}
Cross-correlating the measured voltages from the 512 inputs of CASM-256 to form the full visibility matrix $\mathbf{V}_{ij}$ involves computing the following matrix product (suppressing the frequency dependence): \begin{equation} \mathbf{V}_{ij} = \mathbf{v}_{i}^t\left(\mathbf{v}_{j}^t\right)^\dagger, \end{equation} where $i,j$ label the inputs, $\mathbf{v}_{i}^t$ is a $(N_{\text{ant}}N_{\text{pol}}) \times N_{\text{samp}}$ vector of observed complex voltage samples, and $^\dagger$ denotes the Hermitian conjugate. $\mathbf{V}_{ij}$ is thus a $(N_{\text{ant}}N_{\text{pol}})^2$ Hermitian matrix, of which only the upper (or lower) half needs to be computed.

Correlation is performed as part of the same GPU-based pipeline as the beamforming described above, with the matrix multiplications implemented using CUTLASS. As with the direct beamformer, the matrix multiplications are performed using tensor cores with 8-bit integer input data accumulated to 32-bit integer output.
This configuration achieves approximately 155 TOPS, which is sufficient to allow simultaneous beamforming and correlation.

As discussed in Section \ref{subsec:packet_capture}, the input data are processed in blocks of 2048 samples corresponding to approximately $67\,\mathrm{ms}$.
However, we do not require a new visibility matrix to be written out for every input block --- instead several blocks' worth of visibility data are integrated before being passed on to the visibility processing pipeline. 
The cadence with which visibilities are produced is configurable, but a typical value is approximately once every minute.

\subsection{Single-pulse search} \label{subsec:single_pulse}

We employ a modified version of the GPU-based dedispersion software, 
{\tt hella}, written for the DSA-110 by Vikram Ravi. 
The adopted code has been optimized for the CASM search problem\footnote{https://github.com/Coherent-All-Sky-Monitor/casm-hella}. The pipeline is written in CUDA C++. {\tt hella} first RFI cleans all intensity beams, after which the beams are 
dedispersed at a range of DMs using {\tt dedisp}. The DM/time array is then convolved with a series of boxcar filters and $S/N\geq6$ peaks are identified at different DM, time, beam, and pulse width values. As CASM is concerned with Galactic transients and nearby extragalactic FRBs, our DM range must extend down to nearly zero. We set a maximum 
DM value of 1000\,pc\,cm$^{-3}$, limited by compute. After peak finding, candidates will be sent through a series of clustering, filtering, and ML-based classification algorithms. Dispersed impulsive signals in the far field have a unique response in beam/DM/time/width space, allowing us to identify real events and reject RFI based on the distribution of candidates post dedispersion. FRB candidates that survive the filtering stages will 
trigger a dump of voltage buffer to disk from all core and outrigger antennas, enabling offline interferometric localization. The voltage dumping system 
accounts for the dispersive sweep of the pulse. We plan to save 2-4\,s of data 
at each of the six sub-bands, with the pulse at the center of each window at the central sub-band frequency.

\begin{figure}[htbp]
    \centering
    \includegraphics[width=\textwidth]{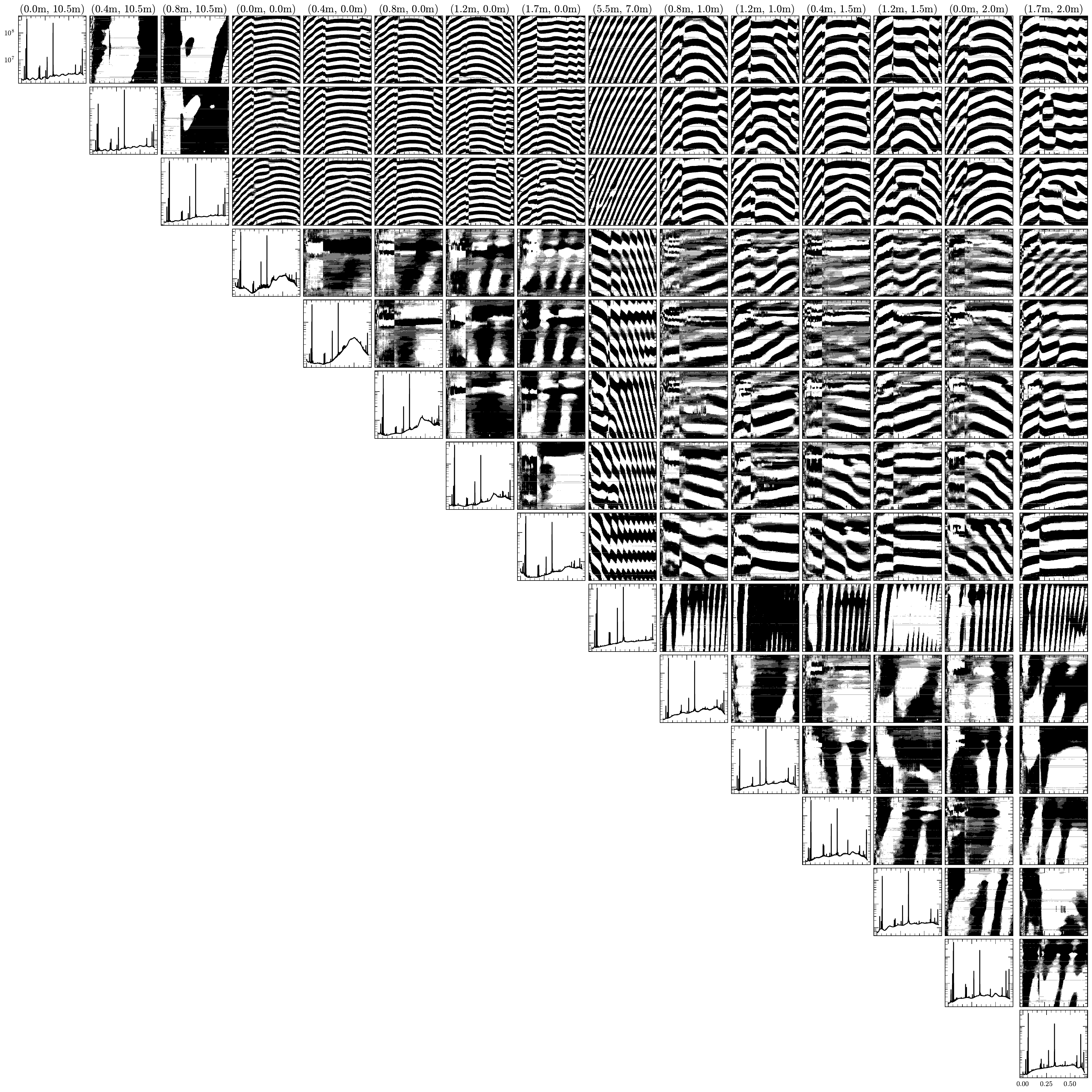}
    \caption{\textbf{The correlation matrix during solar transit.} Each off-diagonal panel corresponds to 5.5 hours of visibility data as a function of frequency (vertical axis) and time. We plot a clipped real-component of $V_{ij}$. 
    The impact of cross-talk can be seen in short ($\lesssim$\,3\,m) baselines. Diagonals show each antenna's unflattened auto-spectrum.}
    \label{fig:xcorr}
\end{figure}

\section{Data from commissioning array}
During an early deployment run in December 2025, 6 planks 
of 6 PCB antennas were built and fastened to the groundscreen. The full analog and digital signal chain has been deployed 
for 18 of these antennas, including the antenna west of the core. We added a single antenna on a separate groundscreen, $\sim$5\,m west of the core, providing a test east-west baseline to the mostly north-south core array. Signals received from the sky 
are amplified by the LNAs, sent over 10--20\,m of coaxial 
cable, digitized and channelized by the SNAP boards, 
sent through the vault's 100\,Gb network switch and put on fiber 
terminating at our GPU server in the correlator room.
Over the past two months, data have been captured in real-time over multiple NICs on our GPU servers and fed to a 128-input correlator. We have established that our pipeline can handle 
data rates above the per-NIC data rate of the final, 256$\times$2\,pol input system. In Figure~\ref{fig:array}, we 
show the arrays first-light fringes for a single baseline 
between the central plank and the extra antenna. The data are
a solar transit on December 13th, 2025. The Sun was low on the sky during this time ($\sim$\,29$^\circ$ above the horizon), making it difficult to interpret the beam shape below our antenna's FWHM. 
Still, per channel S/N is in line with our expectations for a $300-500$\,kJy source at this position in our primary beam. A sub-set of the 128$\times$128 visibility 
matrix is shown in Figure~\ref{fig:xcorr}. These are the channels with full analog chains during that observation.

\begin{figure}[htbp]
    \centering
    \includegraphics[width=0.50\textwidth]{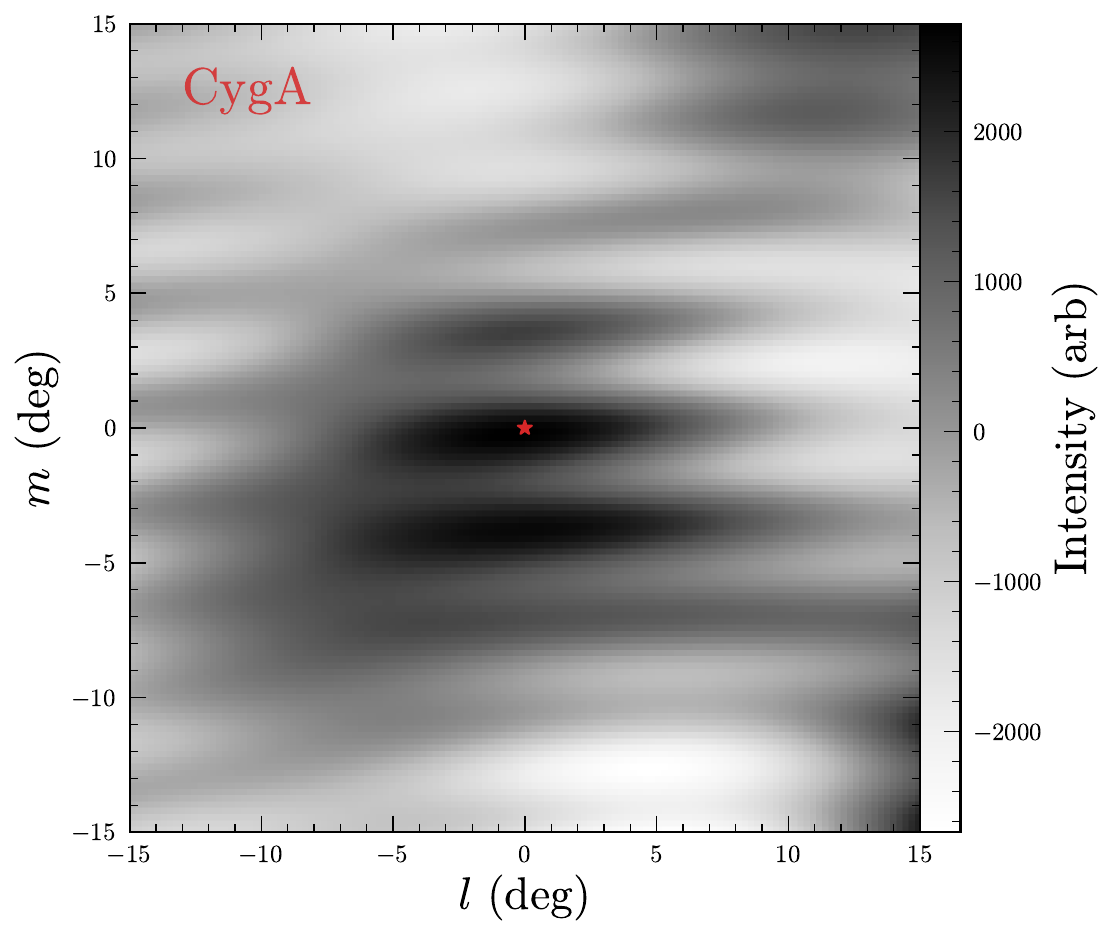}
    \caption{\textbf{A narrow-band (420-423\,MHz) dirty image of Cygnus A.} Visibilities from 15 antennas were imaged shortly before sunrise with phase center at the position of Cyg A. $l$ and $m$ are direction cosines.}
    \label{fig:cyga}
\end{figure}

\subsection{Cross-talk} 

A dense aperture array like CASM will inevitably be impacted by mutual coupling between antennas. The question is, to what extent does it impact our science and can it be calibrated out? CHIME is a densely packed interferometer with four parabolic cylinders, each consisting of a focal line with 256 dual-polarization PCB antennas that share a reflector. Mutual-coupling along the focal line has a significant impact on the CHIME 21\,cm cosmology experiment, which requires precise calibration \citep{Amiri_2022,sanghavi}. Its FRB experiment does not explicitly account for cross-talk, and CHIME/FRB is the most prolific FRB experiment to date \citep{Amiri2018CHIMEFRBSystem, CHIMEFRB2026Catalog2}. Phased-array feeds (PAFs) also 
consist of many nearby antennas. In that case, cross-talk between PAF inputs is solved in a generalized weights matrix and data are corrected during when PAF beams are created.
This is common practice in both communications and radio astronomy applications of PAFs \citep{Voronkov2008PAF, Landon_2010, LI20161065}.

The impact of cross-talk is apparent for intra-plank baselines in early CASM-256 data. Mutual coupling between antennas is likely responsible, rather than signals mixing between channels in our back-end. Fortunately, this coupling appears to be stationary in time, showing up as constant time offsets in the visibilities. 
Visibilities of nearby antennas are not mean-zero on timescales of hours to days. Longer baselines do have zero mean. We are currently investigating if our beamformer weights require a decoupling matrix that modifies the standard beamformer weights matrix. 

Without cross-talk, the observed per-antenna voltage is,

\begin{equation}
    \mathbf{v_{obs}} = \mathbf{G\,a(\hat{s})}s\,+\mathbf{n}
\end{equation}

\noindent where $s$ is a bright point-source sky signal in direction 
$\bf \hat{s}$ (the Sun, e.g.), $\mathbf{a}$ is the steering vector for that direction, and 
$\mathbf{G}$ is a matrix with $N_{ant}$ complex 
gains. The observed visibility matrix is,

\begin{equation}
   \mathbf{V_{obs}} = s^2\,\bf G\,a a^\dagger\,G^\dagger + N
   \label{eq:xtalk}
\end{equation}

\noindent we would then solve for $\bf G$ and apply beamformer weights in the standard way--assuming the Sun is sufficiently bright and compact relative to our beams that $\bf V$ is nearly rank-one. However, if signals are mixed between different antennas, we must 
account for a mutual-coupling/cross-talk matrix $\bf C$.

\begin{equation}
    \mathbf{v_{obs}} = \mathbf{C\,G\,a(\hat{s})}s\,+\mathbf{n}\,\,\,\,\,\mathrm{and} \,\,\,\,\,\mathbf{V_{obs}} = s^2\,\bf C\,G\,a a^\dagger\,G^\dagger\,C^\dagger + N
\end{equation}

Here, $\bf C$ has complex elements $\epsilon_{ij}$ that 
correspond to the component of antenna $j$ measured 
by antenna $i$. By construction, $\epsilon_{ii}=1$ and 
no mutual coupling means $\mathbf{C} = \mathbf{I}_{N_{ant}}$, returning us to Equation~\ref{eq:xtalk}. In most PAF applications, a general calibration solution absorbs the cross-talk: $\bf C$ and $\bf G$ are not solved for separately. Instead, $\bf M\equiv GC$ is determined 
and beamformer weights can be computed as,

\begin{equation}
    \mathbf{w} = \frac{\mathbf{V}_{obs}^{-1}\,\mathbf{a}_{eff}}{\mathbf{a}_{eff} \mathbf{V}_{obs}^{-1}\,\mathbf{a}_{eff}^\dagger}
\end{equation}

\noindent where $\mathbf{a}_{eff}\equiv \bf M\,a$.

\begin{figure}[htbp]
    \centering
    \includegraphics[width=0.65\textwidth]{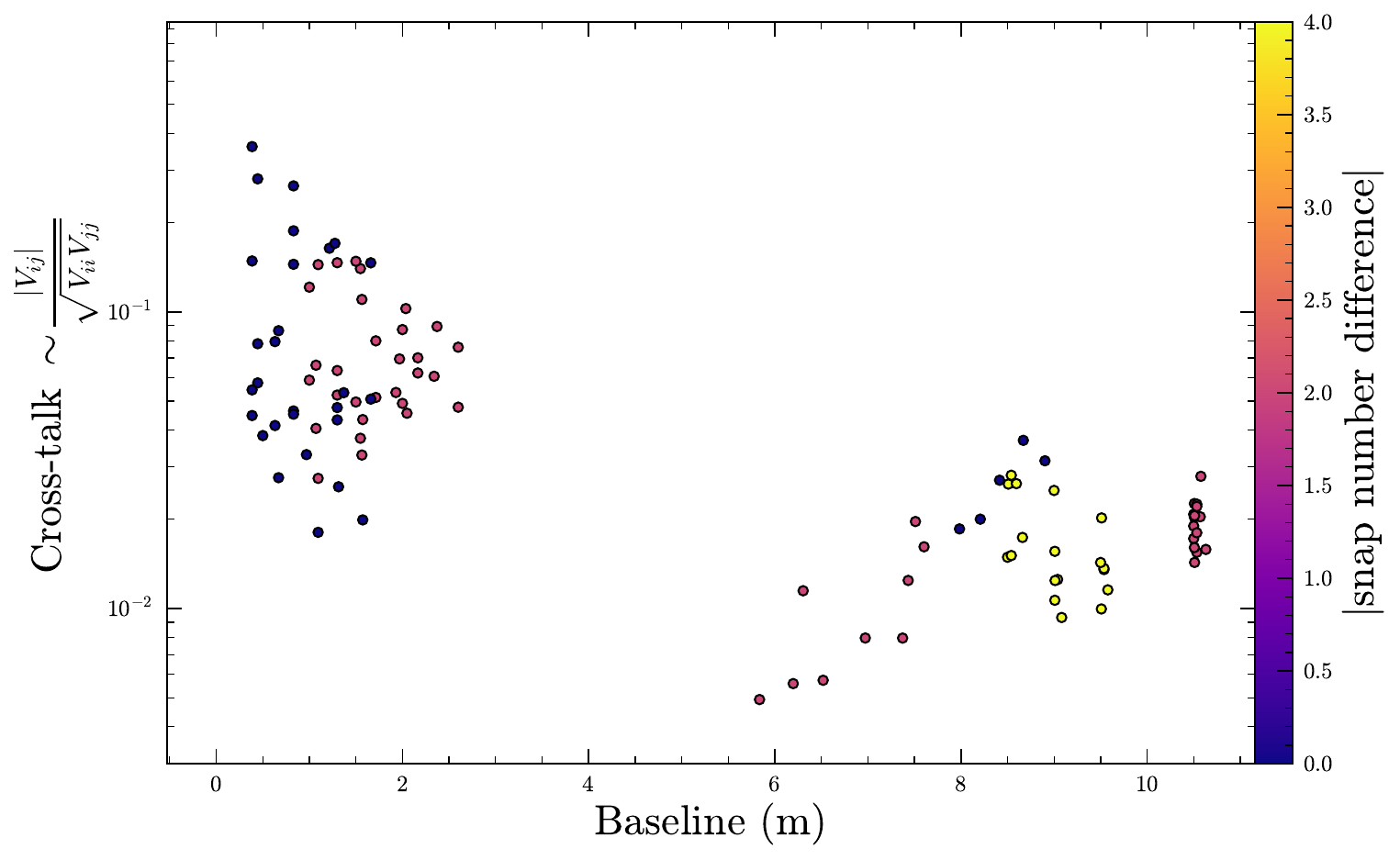}
    \caption{\textbf{An estimate of cross-talk amplitude vs. baseline length for the first 16 CASM antennas}. In the absence of cross-talk, $V_{ij}/\sqrt{V_{ii}V_{jj}}$ ought to have mean zero when there is no bright source in the beam and the amplitude of the $y-$axis would be due to noise and residual sky signal. As expected, cross-talk is largest for adjacent antennas, falling to several percent for baselines longer than 5\,m. Color corresponds to the difference 
    in SNAP board number, meaning blue points share a SNAP board. There may be evidence of digital coupling in the 8--9\,m intra-SNAP baselines in this figure.}
    \label{fig:xtalk}
\end{figure}

For CASM-256, we expect $\mathbf{C}$ to be block-diagonal. 
From early data, we find that antennas separated by 
multiple have little evidence of cross-talk. We estimate the magnitude of cross-talk on CASM-256 
with $\epsilon_{ij} \approx \frac{V_{ij}}{\sqrt{V_{ii}\,V_{jj}}}$ when the Sun is not in our beam. In Figure~\ref{fig:xtalk}, we find that $\epsilon_{ij}\sim2-35\%$ for nearby antennas and 
$\epsilon_{ij}\lesssim3\%$ for antennas separated by more than $5$ meters. 
During nighttime observations $V_{ij}$ 
still includes sky signals, such as diffuse Galactic emission and CygA/CasA, so these $\epsilon$ values can be considered upper-limits. 
In a future work (Sanghavi et al. (in prep)), we will 
investigate the material impact of cross-talk on our science and determine what remedies are needed.

\subsection{Testing the real-time pipeline}

\begin{figure}[htbp]
    \centering
    \includegraphics[width=0.65\textwidth]{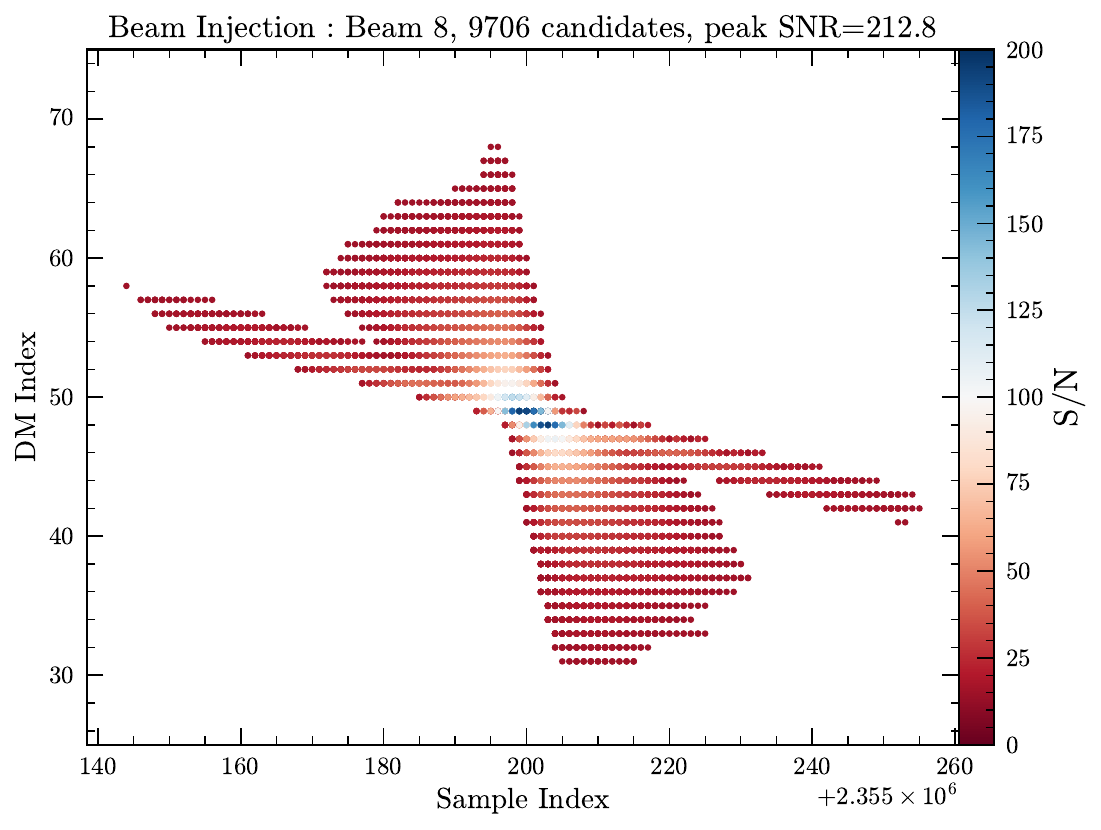}
    \caption{\textbf{Single pulse injection recovered by the real-time pipeline.} We inject simulated FRBs into intensity beams to test the performance of the CASM 
    real-time search. Shown here is an example of a ``bow-tie'' of candidates for a high S/N injection, recovered at the correct DM.}
    \label{fig:injection}
\end{figure}

We have an operational real-time pipeline that captures data from our SNAP boards 
via the 100\,Gb network switch, runs a correlator and beamformer on six sub-bands of 512 channels, searches 
for single pulses, and dumps voltage data when triggered. Beamformer weights 
are generated by fringestopping visibility data and then solving for 
per-antenna complex gains from the Sun with a singular value decomposition (SVD). Calibration weights are applied to the geometric weights matrix for 512 beams, which is then used for real-time beamforming. Sub-band beams with 1\,ms sampling and 30.5\,kHz channels are then cornerturned 
such that three GPUs on each server have one sixth of the beams for all frequencies. 
Our single-pulse search software, {\tt casm\_hella}, dedisperses and finds peaks at different DMs, times, pulse widths, and beams. We inject simulated FRBs into the real-time pipeline. An example of a recovered injection is shown in Figure~\ref{fig:injection}. Clustering and classification of candidates will 
then decide if voltage data for a given trigger is to be dumped. At this point in 
the commissioning phase, voltage dumps are triggered manually.

\subsection{Radio frequency interference}
The terrestrial RFI environment at OVRO between 
200-600 MHz is relatively clean (see Figure~\ref{fig:rate}).
While RFI from the horizon has been characterized, early commissioning data suggests that the most pernicious 
source of RFI will be from the sky. The proliferation 
of satellite constellations and direct-to-cell signals 
present a significant threat to all of radio astronomy, including all-sky telescopes like CASM and BURSTT. 
The Mobile User Objective System (MUOS) is the U.S. Navy’s narrowband military satellite communications network, consisting of five geosynchronous satellites and operating at UHF. Three of these are above the horizon at OVRO. We detect 
the MUOS downlink (360-380\,MHz) at the bottom of our band. The ground-based uplink (300–320\,MHz) was 
found to alias into our digital band 
before a high-pass filter 
was employed to cut off signals at $\leq400$\,MHz. We have developed software to monitor potential aerial RFI sources in our beam. We have a tool for near real-time recording and visualization of satellites and aircraft above the horizon at an observatory location\footnote{https://github.com/Coherent-All-Sky-Monitor/casm\_sky}. As with the Sun, strong RFI sources can potentially be 
``nulled'' in our time-dependent beamformer weights, provided we know 
where they are at all times.

Our real-time pipeline mitigates RFI at multiple stages. First, we mask
bad channels in our F-engine by applying weight zero to perennially 
contaminated frequency channels. After the beamformer cornerturn, 
all Stokes I beams are cleaned using standard iterative thresholding techniques. This is done on a per-beam basis. 
These include bandpass flattening, DM zero subtraction, and spectral flagging with a series of smoothing lengths. 
After dedispersion and peak finding, false-positive candidates produced by non-Gaussian near-field RFI sources are clustered, classified, and discarded. True impulse sky signals such as pulsars and FRBs ought to 
look like a PSF on the sky and like a bow-tie in DM/time space. Horizon RFI will occupy many beams and satellite RFI will show up at many DMs. 

\section{A future CASM with $>10^{4}$ antennas}
Dense aperture arrays like CASM are highly scalable because they 
are composed of many identical low-cost parts whose assembly can 
be outsourced to industry. More importantly, the enormous data rates produced by large-$N$ radio arrays can be processed on increasingly powerful GPUs. The per-unit cost of a CASM-256 PCB antenna is roughly \$50. The LNA and back-end board are also below \$100 per board and chassis. The hardware budget of 
CASM-256 is dominated by digitization, channelization, and 
the GPU servers that process streaming data. Raw data rates are linear in the number of antennas while fast transient detection rate scales as $N_{ant}^{1.5}$. Mapping speed scales as $N_{ant}^2$. Therefore, 
one not only gets more science from larger arrays, but significantly more science-per-dollar. The railroad track on 
which CASM-256 is being built could, in principle, support 
up to 1536 antennas in a $6\times256$ grid. The detection rate of 
such an array would exceed $10^3$ per year.
We are considering modest modifications to the current design to propose expanding along southward along rail.

If the analog signal chain can be optimized for performance and 
the per-channel digital backend cost can be minimized, an array 
with tens of thousands of antennas could be built at the scale of
a large NSF Mid-scale Research Infrastructure-1 or small MSRI-2---comparable to a multi-fiber O/IR spectrograph. An array with tens of thousands of antennas would require a new site and significant design effort.

A future CASM with, e.g., 32,786 antennas, would detect $10^6$ FRBs in a 5-year survey. Such a sample would produce $\mathcal{O}(100)$ strongly lensed FRBs, potentially enabling ambitious applications like measuring the Universe's expansion in real-time \citep{Wucknitz2021}. The FRB DMs would make tomographic maps of the cosmic baryons at $0<z<1.5$, answering fundamental questions in galaxy formation, and the impact of feedback on precision cosmology. CASM-32k would find all repeating Northern FRBs within 1\,Gpc, all Galactic analogs, and radio pulsars from external galaxies above 0.04\,Jy\,ms. Digital beams could track and time pulsars nearly continuously. With improved analog design and an optimal antenna configuration, CASM-32k could be optimized for 21\,cm cosmology at $1.5\lesssim z\lesssim6$, though this would require a major RF design effort. If the array were designed for imaging, it would have a higher survey speed than SKA1-mid. 
This is an incomplete list of the high-impact science one 
could do with a sensitive all-sky radio telescope. We briefly consider the design and cost of such an instrument.

Without further optimization, the CASM-256 design could be scaled to an 
array with 32\,k antennas with a hardware cost of roughly $\$50$\,M in 2025 USD. This would be suboptimal, as CASM-256 uses $\sim$\,15 year old technology for digitization and channelization (SNAP boards). SNAPs were chosen for CASM-256 as each board has 12 ADC inputs that can operate at 250\,MS/s and because our group has experience with SNAP boards. The F-engine is a major cost driver of ultra-large $N$ arrays, so a custom board with 32 to 64 inexpensive ADCs ($\sim$500\,MS/s) would be ideal. We consider this a better alternative to multiplexing many antennas on RFSoC boards with fast digitizers (e.g., RFSoC 4×2 with 5GS/s ADCs). Channelization could be done on GPUs instead of FPGAs. There are now high-performance, operational GPU-based channelization systems \citep{merry2023}. In that case, sampling 
would happen near the antennas and packetized digital baseband data would be sent to a GPU cluster over fiber, where they would be channelized, beamformed, etc. all in one system. The digital front-end board would need a small FPGA handling the ADC samples, streaming, and synchronization. A design study will need to be done to 
assess the tradeoffs between a GPU-based F-engine and the standard FPGA solution. The current CASM design requires roughly 10\,km of coaxial cable to bring RF from the antenna/LNA to our F-engine. A scaled version of CASM must digitize near the antennas and put signals onto fiber early in the front-end chain. 

On the analog side, CASM-256 uses connectorized LNAs matched to 50\,$\Omega$ and a separate board 
for bandpass filtering, power, and pre-ADC amplification.  A more performative,
streamlined design would jointly optimize an antenna with embedded 
analog electronics. The $N_{ant}^2$ compute requirements of cross-correlation and beamforming can be circumvented with techniques 
like FFT interferometry ($\mathcal{O}(N_{ant}\log N_{ant})$) at the cost of redundant or semi-regular antenna layout. Low-bit depth 
operations on NVIDIA tensor cores can massively accelerate the software pipeline both in compute and memory requirements. Ultra high-density 
compute solutions such as the NVL72 systems 
could capture the full data rate of CASM-32k on a single-rack. 
Indeed, the upcoming Deep Synoptic Array (DSA) \citep{hallinan2019} is considering the NVL technology for its Radio Camera and Chronoscope instruments. These turn-key racks solve the antenna/frequency/beam corner turn problem in time-domain radio astronomy by having fast interconnect across GPUs. The DSA data rate is comparable to a CASM-32k with 
100\,MHz of bandwidth, but DSA requires many more beams due to its $\sim$\,20\,km maximum baseline--even accounting for the large FoV of an aperture array. CASM's dense antenna configuration means fewer beams/pixels need to be created, and less data must be searched. 

If antennas for CASM-32k were as tightly packed as CASM-256 
and laid out on a square grid, its footprint would be just 90$\times$90\,m. But antennas would be inaccessible for installation and maintenance, and severe cross-talk could limit sensitivity. Such a telescope would also have poor angular resolution, impacting non-FRB science. 
Instead, tiles of $16\times16$ antennas (for example) could be distributed on uniform grid points, allowing for FFT-beamforming.
A sparser array would enable better imaging performance and higher angular resolution at the cost of number of beams and therefore compute. Discovering the optimal antenna configuration requires optimizing for the PSF, regularized by key science drivers and compute tradeoffs in FFT beamforming/correlation vs. brute-force methods. This can be done with a differentiable computation graph on which gradient-based optimization could be run--similar to how the pseudorandom array configuration was discovered for the 1650 DSA antennas.

We estimate that the per-channel analog and digital cost could be brought down from $\$$1\,k to $\lesssim$\,\$500, 
i.e. a total hardware cost of \$20\,M for a CASM-32k. Projects far above this scale are common in industry. Solar farms are routinely built with $10^5-10^7$ 
identical solar panels, densely packed. More relevant to CASM, SKA-low 
will have $\sim10^5$ individual antennas\footnote{https://www.skao.int/en/explore/telescopes/ska-low} organized in 
tiles of 256 log-periodic dipoles in a pseudorandom configuration that are beamformed to produce ``station beams''. SKA-low will not be an all-sky monitor because 
the effective instantaneous FoV of a station beam is degree-scale. This choice was made to maximize a wide range of science cases given data rate constraints.
We are suggesting a continuous all-sky telescope that processes the full $10^4$\,deg$^2$ FoV, maximizing survey speed and fast transient science rate per dollar. 

\section{conclusions}

We have described the 256-antenna Coherent All-Sky Monitor 
and early stages of its deployment at the Owens Valley 
Radio Observatory (OVRO). The first 37 antennas have been deployed on a metal groundscreen in plastic radomes. Two dozen of those antennas have fully-connected analog chains that are sampled by 
12-input SNAP boards and output over fiber to correlator nodes via a 100\,Gbe switch. We plan to deploy the remaining 256 analog chains in the next six months. An end-to-end software pipeline is running, including packet capture, beamforming, cross-correlation, and single-pulse searching. We have tested our data by beamforming on, and imaging the Sun and CygA. We use the Sun for beamformer weight calibration. Per-baseline sensitivity is roughly inline with expectations, though we have not rigorously characterized the SEFD of the phased array. Outrigger stations for FRB localization are under design and will likely deploy in two stages: on-site ($\sim$5'') and off-site (sub-arcsecond). Finally, we consider CASM a pathfinder instrument for a much larger array that could detect one million FRBs, in addition to many other science cases.

\begin{acknowledgements}                                                                   
We thank the Caltech's Owens Valley Radio Observatory (OVRO) for              
hosting CASM-256 and OVRO staff for supporting the project. 
We are grateful to the Mt.~Cuba Astronomical              
Foundation for funding the CASM hardware, as well as the Harvard College Observatory 
for support. L.~Connor acknowledges support from
the National Science Foundation under grant No. 2537086. We also thank Keith Bannister, Aaron Parsons, and Dan Werthimer for helpful discussions.                           
\end{acknowledgements}  

\bibliography{PASPsample701}{}
\bibliographystyle{aasjournalv7}



\end{document}